%% file: ggc.tex
\documentclass[useAMS,usenatbib,iop,numberedappendix]{mn2e} 

\usepackage{mathptmx}
\usepackage{amsmath,amsfonts,amssymb}
\usepackage{mathrsfs}
\usepackage{makeidx}
\usepackage{graphicx}
\usepackage{epstopdf}
\usepackage[colorlinks=True]{hyperref}
\usepackage{multirow}
\usepackage{rotating}
\usepackage{color}
\usepackage{tablefootnote}
\usepackage{lscape}
\usepackage{subcaption}
\usepackage{tablefootnote}
\usepackage[T1]{fontenc}
\usepackage{aecompl}
\usepackage[normalem]{ulem}
\usepackage{siunitx}

\definecolor{inon}{rgb}{1.00,0.27,0.00}

\input{authorlist}

\newcommand{\LCDM}{\ensuremath{\Lambda\textrm{CDM}}}
\newcommand{\OmegaM}{\ensuremath{\Omega_{\mathrm{M}}}}

\newcommand{\Omegab}{\ensuremath{\Omega_{\mathrm{b}}}}
\newcommand{\Hnow}{\ensuremath{H_{0}}}
\newcommand{\seight}{\ensuremath{\sigma_{8}}}
\newcommand{\Msun}{\ensuremath{\mathrm{M}_{\odot}}}

\newcommand{\Rfiveoo}{\ensuremath{R_{500}}}
\newcommand{\Mfiveoo}{\ensuremath{M_{500}}}

\newcommand{\Mfiveootilde}{\ensuremath{\tilde{M}_{500}}}
\newcommand{\redshift}{\ensuremath{z}}
\newcommand{\dif}{\ensuremath{\mathrm{d}}}

\newcommand{\rhocrit}{\ensuremath{\rho_{\mathrm{c}}}}

\newcommand{\PLANCK}{\emph{Planck}}
\newcommand{\WMAP}{\emph{WMAP}}

\newcommand{\Pmm}{\ensuremath{P_{\mathrm{mm}}}}
\newcommand{\ximm}{\ensuremath{\xi_{\mathrm{mm}}}}
\newcommand{\xihh}{\ensuremath{\xi_{\mathrm{hh}}}}
\newcommand{\xihoht}{\ensuremath{\xi_{\mathrm{X}\mathrm{Y}}}}
\newcommand{\cmr}{\ensuremath{\mathbf{r}}}
\newcommand{\cms}{\ensuremath{\mathbf{s}}}
\newcommand{\kk}{\ensuremath{\mathbf{k}}}
\newcommand{\biash}{\ensuremath{b_{\mathrm{h}}}}
\newcommand{\biasho}{\ensuremath{b_{\mathrm{X}}}}
\newcommand{\biasht}{\ensuremath{b_{\mathrm{Y}}}}

\newcommand{\btinker}{\ensuremath{b_{\mathrm{T10}}}}
\newcommand{\bproj}{\ensuremath{b^{\mathrm{proj}}}}
\newcommand{\bnonproj}{\ensuremath{b^{\mathrm{non-proj}}}}
\newcommand{\bmodel}{\ensuremath{b_{\mathrm{model}}}}
\newcommand{\growthrte}{\ensuremath{\mathfrak{f_{\mathrm{m}}}}}

\newcommand{\zcl}{\ensuremath{z_{\mathrm{cl}}}}

\newcommand{\MPIV}{\ensuremath{M_{\mathrm{piv}}}}
\newcommand{\ZPIV}{\ensuremath{z_{\mathrm{piv}}}}

\newcommand{\Arich}{\ensuremath{A_{N}}}
\newcommand{\Brich}{\ensuremath{B_{N}}}
\newcommand{\Crich}{\ensuremath{C_{N}}}
\newcommand{\Drich}{\ensuremath{\sigma_{N}}}

\newcommand{\rtm}{\ensuremath{N}--\ensuremath{M}}

\newcommand{\xicamira}{\ensuremath{\xi_{\mathrm{cc}}}}
\newcommand{\xicmass}{\ensuremath{\xi_{\mathrm{gg}}}}
\newcommand{\xicamiracmass}{\ensuremath{\xi_{\mathrm{c}\mathrm{g}}}}

\newcommand{\rich}{\ensuremath{N}}

\newcommand{\percent}{\ensuremath{\%}}

\newcommand{\appropto}{\mathrel{\vcenter{
  \offinterlineskip\halign{\hfil$##$\cr
    \propto\cr\noalign{\kern2pt}\sim\cr\noalign{\kern-2pt}}}}}

\DeclareMathOperator\erf{erf}

\title[Cluster Clustering]{
A clustering-based self-calibration of the richness-to-mass relation of 
CAMIRA galaxy clusters out to $\redshift\approx1.1$ in the Hyper Suprime-Cam survey
}

\hypersetup{linkcolor=blue,filecolor=magenta,citecolor=cyan}

\begin{document}
\pdfpageheight 11.7in
\pdfpagewidth 8.3in

%
%

\maketitle 

%
%

\begin{abstract}
We perform a self-calibration of the richness-to-mass (\rtm) relation of CAMIRA galaxy clusters with richness $\rich\geq15$ at redshift $0.2\leq\redshift<1.1$ by modeling redshift-space two-point correlation functions.
These correlation functions are
the auto-correlation function \xicamira\ of CAMIRA clusters, the auto-correlation function \xicmass\ of the CMASS galaxies spectroscopically observed in the BOSS survey, and 
the cross-correlation function \xicamiracmass\ between these two samples.
We focus on constraining the normalization \Arich\ of the \rtm\ relation 
with
a forward-modeling approach, carefully accounting for the redshift-space distortion, the Finger-of-God effect, and the uncertainty in photometric redshifts of CAMIRA clusters.
The modeling also takes into account the projection effect on the halo bias of CAMIRA clusters.
The parameter constraints are shown to be unbiased according to validation tests using a large set of mock catalogs constructed from N-body simulations.
At the pivotal mass $\Mfiveoo=10^{14}h^{-1}\Msun$ and the pivotal redshift $\ZPIV = 0.6$, the resulting normalization \Arich\  is constrained as
$13.8^{+5.8}_{-4.2}$, 
$13.2^{+3.4}_{-2.7}$, and
$11.9^{+3.0}_{-1.9}$
by modeling $\xicamira$, $\xicamira+\xicamiracmass$, and $\xicamira+\xicamiracmass+\xicmass$, with average uncertainties at levels of
$36\percent$, 
$23\percent$, and
$21\percent$, respectively.
We find that the resulting \Arich\ is statistically consistent with those independently obtained from weak-lensing magnification and from a joint analysis of shear and cluster abundance, with a preference for a lower value at a level of $\lesssim1.9\sigma$.
This implies that the absolute mass scale of CAMIRA clusters inferred from clustering is mildly higher than those from the independent methods.
We discuss the impact of the selection bias introduced by the cluster finding algorithm, which is suggested to be a subdominant factor in this work. 
\end{abstract}

%
%


%
%

\section{Introduction}
\label{sec:introduction}

Galaxy clusters are powerful cosmological tools because they provide a representative view of large-scale structures of the Universe.
Therefore, galaxy clusters enable independent tests to examine viable cosmological models with strong constraints on fundamental properties, such as the degree of inhomogeneity in cosmic density fields and the equation of state of dark energy \citep[e.g.,][]{wang1998,holder01b}.
With the progress in utilizing the technique of weak gravitational lensing to calibrate the mass of clusters \citep{umetsu14,vonderlinden14b,vonderlinden14a,hoekstra15,schrabback18,dietrich19,mcclintock19}, there have been 
successful demonstrations of
constraining cosmology by using the abundance of galaxy clusters identified in the millimeter wavelength \citep{PlanckCollaboration2015b,bocquet15,deHaan16,bocquet19}, in X-rays \citep{mantz15}, and in the optical \citep{costanzi18}.
The recent development of cluster cosmology has promised a competitive cosmological tool that is complementary to and independent of other probes, especially 
those relying on the temperature anisotropy of Cosmic Microwave Background (CMB).

Despite the success of constraining cosmology by using cluster abundance, there have been relatively less efforts in utilizing the 
clustering of galaxy clusters in a cosmological analysis.
This was mainly due to the fact that galaxy clusters, as peaks of cosmic density fields, are rare, which inevitably results in insufficient constraining power in terms of two-point or higher-order statistics.
For example, the baryon acoustic oscillation (BAO) signature of galaxy clusters was only marginally detected at a level of $\approx2\sigma$ by using the largest cluster catalog 
available a decade ago \citep{estrada09,hutsi10}.
This situation of lacking a sizable sample of clusters will be rapidly improved with upcoming large and deep surveys, such as the 
Legacy Survey of Space and Time (LSST) carried out by the Vera C. Rubin Observatory \citep{ivezic08}, the \textit{Euclid} mission \citep{laureijs11}, and the \textit{eROSITA} X-ray all-sky survey \citep{merloni12}.
Thus, it is essential and imminent to study the clustering of galaxy clusters, paving a way for upcoming data sets.

Apart from the signature of BAO, few pilot studies were carried out to measure the large-scale clustering of galaxy clusters identified in the existing or ongoing surveys, in which some of them were further used to infer cosmology:
The correlation functions of optically selected clusters were measured and compared to simulations in \cite{bahcall2003}.
\citet{collins2000} measured the correlation function of a X-ray flux-limited sample of $\approx450$ clusters at redshift $\redshift\lesssim0.3$, for which the measurements together with those of cluster abundance were used to constrain cosmological parameters \citep{schuecker03}.
Later, \cite{mana13} demonstrated that a joint analysis of cluster abundance and clustering could significantly improve the constraints on the amplitude of the density fluctuation \seight\ and the matter density \OmegaM\ by an amount of $\approx50\percent$.

Meanwhile, the development of the advanced cluster finding algorithm, \texttt{redMaPPer} \citep{rykoff14}, significantly improved the size and quality of cluster samples constructed in the Sloan Digital Sky Survey \citep[hereafter SDSS;][]{york2000}, which further bolstered the study of cluster clustering.
By using a sample of $\approx120k$ redMaPPer clusters at redshift $\redshift\lesssim0.3$, \cite{sereno15} presented a joint analysis of weak lensing and clustering, for the first time, on the cluster-scale.
A similar work was done in \cite{jimeno15}, where they secured the redshift determination of  redMaPPer clusters by utilizing the spectra from the Baryon Oscillation Spectroscopic Survey \citep[hereafter BOSS;][]{dawson13} and combined the measurements of cluster abundance and clustering to infer cosmology.
\cite{baxter16} constrained the observable-to-mass scaling relation of redMaPPer clusters based on angular clustering alone.
In addition, the angular cross-correlation between redMaPPer clusters and a sample of photometrically selected galaxies was also studied in \cite{paech17}.

To achieve the goal of precision cosmology, it is absolutely necessary to combine the information from both cluster abundance and clustering to tighten the constraint on parameters in order to discover possible failures of the concordance  \LCDM\ cosmological model, and/or to identify new systematics to explain tensions among different probes.
For example, a combined analysis of cluster abundance and clustering sheds light on properties of cosmological neutrinos \citep{marulli11,emami17}.
Moreover, there are distinct advantages in using galaxy clusters as tracers of large-scale structures.
As galaxy clusters are the most massive and gravitationally dominated objects in the Universe, their halo bias, which describes the strength of clustering on large scales with respect to the underlying dark matter, is less sensitive to baryonic properties and environmental effects, which are often referred to as the ``astrophysical bias''. 
This results in a cleaner connection between the underlying matter density and galaxy clusters, of which the halo bias is relatively easier to be characterized through N-body simulations \citep[e.g.,][]{tinker10} than on the galaxy-scale.
Meanwhile, it is of critical importance to combine different surveys, especially those observed spectroscopically.
This is because the clustering strength of a photometrically selected sample on small scales would significantly diminish due to the uncertainty of redshift estimates \citep{sereno15}.
Thus, the inclusion of spectroscopic surveys would significantly improve the accuracy and precision of clustering measurements \citep[e.g., as done in][]{jimeno15}.

In this work, we aim to study the clustering properties of the galaxy clusters optically selected in the Hyper Suprime-Cam (HSC) survey \citep{aihara18a} and their cross-correlation with the CMASS galaxies, which are spectroscopically observed in the BOSS (see Section~\ref{sec:cmass} for more details).
Specifically, we will perform a self-calibration of the observable-to-mass relation based on these clustering measurements alone \citep[see also][]{majumdar03,lima04,hu06,holder06}.
As the deepest optical imaging survey at the achieved area to date, the combination of the depth and area of the HSC survey enables a construction of a sizable sample of clusters for studying
their clustering properties out to high redshift ($\redshift\approx1.1$), for the first time.
Although the clustering signal of galaxy clusters detected in the HSC survey is distorted on small scales because of lacking secure redshifts, the precision of their clustering measurements is significantly improved by cross-correlating with the sample of CMASS galaxies.
The uniqueness of this work is that we perform the mass calibration of galaxy clusters based on halo clustering alone by using a joint data set of the largest cluster sample out to high redshift ($\redshift\approx1.1$) to date and the spectroscopic sample in the common footprint of the BOSS.
It is worth mentioning that a similar analysis is difficult to be achieved in the southern hemisphere using the Dark Energy Survey \citep{des05,des16}, due to the lack of a large spectroscopic sample in a common footprint.
With the upcoming era of large spectroscopic surveys, such as the Dark Energy Spectroscopic Instrument \citep{desi2006} and the Subaru Prime Focus Spectrograph surveys \citep{takada14}, the synergy with imaging and spectroscopic surveys will be common and essential to study galaxy clusters.
In this regard, this work serves a pilot study in this topic.

This paper is organized as follows.
In Section~\ref{sec:basics}, a brief overview of structure formation in the context of halo clustering is provided.
The data products used in this work are described in Section~\ref{sec:data}.
The detailed methodology used to measure the correlation functions are presented in Section~\ref{sec:measurements}.
The modeling of these correlation functions is presented in Section~\ref{sec:modeling}.
We discuss the results in Section~\ref{sec:results}.
The discussions of the selection bias are in Section~\ref{sec:selection}.
The conclusions are given in Section~\ref{sec:conclusions}.
Throughout this paper, we assume a flat \LCDM\ cosmology with
$\OmegaM=0.3$,
the mean baryon density $\Omegab=0.05$,
the Hubble expansion rate $\Hnow = h\times100$\,km\,s\,$^{-1}$\,Mpc$^{-1}$ with
$h=0.7$, $\seight = 0.8$, and 
the spectral index of the primordial power spectrum $n_{\mathrm{s}}=0.95$.
The mass $\Mfiveoo$ of a cluster is defined by a sphere with the radius $\Rfiveoo$, in which the enclosed mass density is equal to $500$ times the critical density $\rhocrit(\redshift)$ of the Universe at the cluster redshift.
Unless otherwise stated, all quoted errors represent $68\percent$ confidence levels (i.e., $1\sigma$). 
The notation $\mathcal{N}(x,y^2)$ ($\mathcal{U}(x,y)$) stands for a normal distribution with the mean $x$ and the standard deviation $y$ (a uniform distribution between $x$ and $y$). 

%
%

\section{Theory}
\label{sec:basics}

An overview of structure formation in the context of halo clustering is given in this section.
We refer interested readers to \cite{mandelbaum13} and \cite{okumura16} for more details.

The two-point statistics of matter distributions is one of the most straightforward ways to describe cosmic structures.
For instance, the correlation function $\ximm(\cmr)$ of matter, which is an inverse Hankel
transform of the matter power spectrum $\Pmm(\kk)$ in the Fourier-space, describes the excess of matter density fields separated by a distance \cmr\ in the comoving coordinate with respect to a random distribution.
That is,
\begin{equation}
\label{eq:def_ximm}
\ximm(\cmr) \equiv \left\langle\delta_{\mathrm{m}}(\mathbf{x})\delta_{\mathrm{m}}(\mathbf{x} + \cmr)\right\rangle \, ,
\end{equation}
where $\delta_{\mathrm{m}}(\mathbf{x})$ is the matter overdensity at $\mathbf{x}$.
The bracket $\left\langle\right\rangle$ stands for the ensemble average over $\mathbf{x}$.

Halos form via gravitational collapse and result in biased tracers of the overall density field.
In the linear perturbation theory, the overdensity $\delta_{\mathrm{h}}$ of a halo population can be linked to the matter overdensity by a halo bias \biash\ as follows,
\begin{equation}
\label{eq:halobias}
\delta_{\mathrm{h}}(\mathbf{x}) = \biash \delta_{\mathrm{m}}(\mathbf{x}) \, ,
\end{equation}
such that the clustering strength of halos reads
\begin{equation}
\label{eq:halobias2}
\xihh(\cmr) = \biash^2 \ximm(\cmr) \, ,
\end{equation}
where \xihh\ is the correlation (or auto-correlation) function of halos, and the halo bias \biash\ mainly depends on the halo mass and redshift.
As an analogy to the auto-correlation function, the cross-correlation function between two populations of halos is expressed as
\begin{equation}
\label{eq:cross-correlation}
\xihoht(\cmr) = \biasho\biasht \ximm(\cmr) \, ,
\end{equation}
where \biasho\ and \biasht\ are the halo bias of the halo population $\mathrm{X}$ and $\mathrm{Y}$, respectively.
In a regime where the linear perturbation theory fails, e.g., on small scales, the halo bias \biash\ could be scale-dependent.
If the halo bias is known, one can determine the correlation function $\xihh$ of a halo population to further unveil the underlying matter distribution.

However, it is challenging to accurately determine three-dimensional correlation functions in observations, because the line-of-sight distance is unknown and must be inferred from observables.
In the context of redshift surveys, the line-of-sight distance to each object is usually inferred from the observed redshift $\redshift_{\mathrm{obs}}$.
Using the inferred distance, the resulting correlation function $\xihh(\cms)$ in the ``redshift-space'', denoted as \cms, is modulated with respect to that in the real-space $\xihh(\cmr)$.
This is because the observed redshift $\redshift_{\mathrm{obs}}$ is deviated from the cosmological redshift $\redshift_{\mathrm{c}}$ due to the presence of the peculiar velocity $v_{\mathrm{pec}}$ of halos and the measurement uncertainty $\Delta\redshift$:
\begin{equation}
\label{eq:redshift}
\redshift_{\mathrm{obs}} =\redshift_{\mathrm{c}} + \frac{v_{\mathrm{pec},\parallel}}{c}\left(1 + \redshift_{\mathrm{c}} \right) + \Delta{\redshift} \, ,
\end{equation}
where the subscript
$\parallel$ denotes the component along the line of sight, and $c$ is the speed of light.

On large scales, halos are experiencing a coherent movement toward the potential center of cosmic structures, as a result of gravitational collapse.
This leads to a squash in the distribution of the line-of-sight distance that is inferred by the observed redshift.
Consequently, the redshift-space correlation function is distorted, as known as the redshift-space distortion \cite[RSD;][or the Kaiser effect]{kaiser1987}.
On small scales, halos act as particles with a random motion due to the presence of peculiar velocity, resulting in a stretch in the distance distribution along the line of sight.
This is a nonlinear RSD effect known as the Fingers-of-God (FoG) effect \citep{jackson1972}.

It is important to note that measurement uncertainties of redshift play an important role in determining redshift-space correlation functions.
In an imaging survey, as used in this work, the redshift is usually estimated by the photometry redshift (or photo-\redshift) with a typical uncertainty $\Delta\redshift$.
Because the line-of-sight comoving distance to a halo at redshift $\redshift$ is 
\[
D_{\mathrm{LoS}} = \int_{0}^{z}\frac{c\dif\redshift^{\prime}}{H(\redshift^{\prime})} \, ,
\]
where $H(\redshift)$ is the Hubble constant, 
a dispersion $\sigma_{\Delta\redshift}$ in the redshift uncertainty would result in a characteristic scale,
\begin{equation}
\label{eq:sigmascale}
\sigma_{\mathrm{LoS}} = \frac{c\sigma_{\Delta\redshift}}{H(\redshift)} \, ,
\end{equation}
such that the line-of-sight clustering signature is significantly smeared out on the scale $\lesssim\sigma_{\mathrm{LoS}}$.
Taking a typical value of $\sigma_{\Delta\redshift}=0.01$  for optically selected clusters, this corresponds to $\sigma_{\mathrm{LoS}}\approx20$~Mpc$/h$ at $\redshift\approx0.3$.
That is, the power spectrum would be largely suppressed at $k\gtrsim1/\sigma_{\mathrm{LoS}}$ due to the photo-\redshift\ uncertainty.
Moreover, this effect is significantly larger than the FoG effect: 
the typical line-of-sight velocity dispersion $\sigma_{v}$ for halos is at the order of $\approx300$~km/sec, which only leads to $\approx\sigma_{v}(1 + z)/c\approx0.0013$ at $\redshift\approx0.3$, i.e., a factor of $\approx10$ smaller than the dispersion of the photo-z uncertainty.
Therefore, the uncertainty of photo-\redshift\ is the dominant factor over the peculiar velocity of halos in determining redshift-space correlation functions in imaging surveys and needs to be modeled \citep{sereno15}.
We refer readers to Section~\ref{sec:modeling} for the detailed modeling of observed redshift-space correlation functions.

In this work, we measure (1) the correlation function of the cluster sample in the HSC survey, (2) the correlation function of CMASS galaxies, which are spectroscopically observed in the BOSS, and (3) their cross-correlation function. 
The goal is to calibrate the observable-to-mass relation, i.e., the richness-to-mass relation, of the galaxy clusters detected in the HSC survey by using these clustering measurements in a joint analysis.

%
%

%
\begin{figure}
\centering
\resizebox{0.5\textwidth}{!}{
\includegraphics[scale=1]{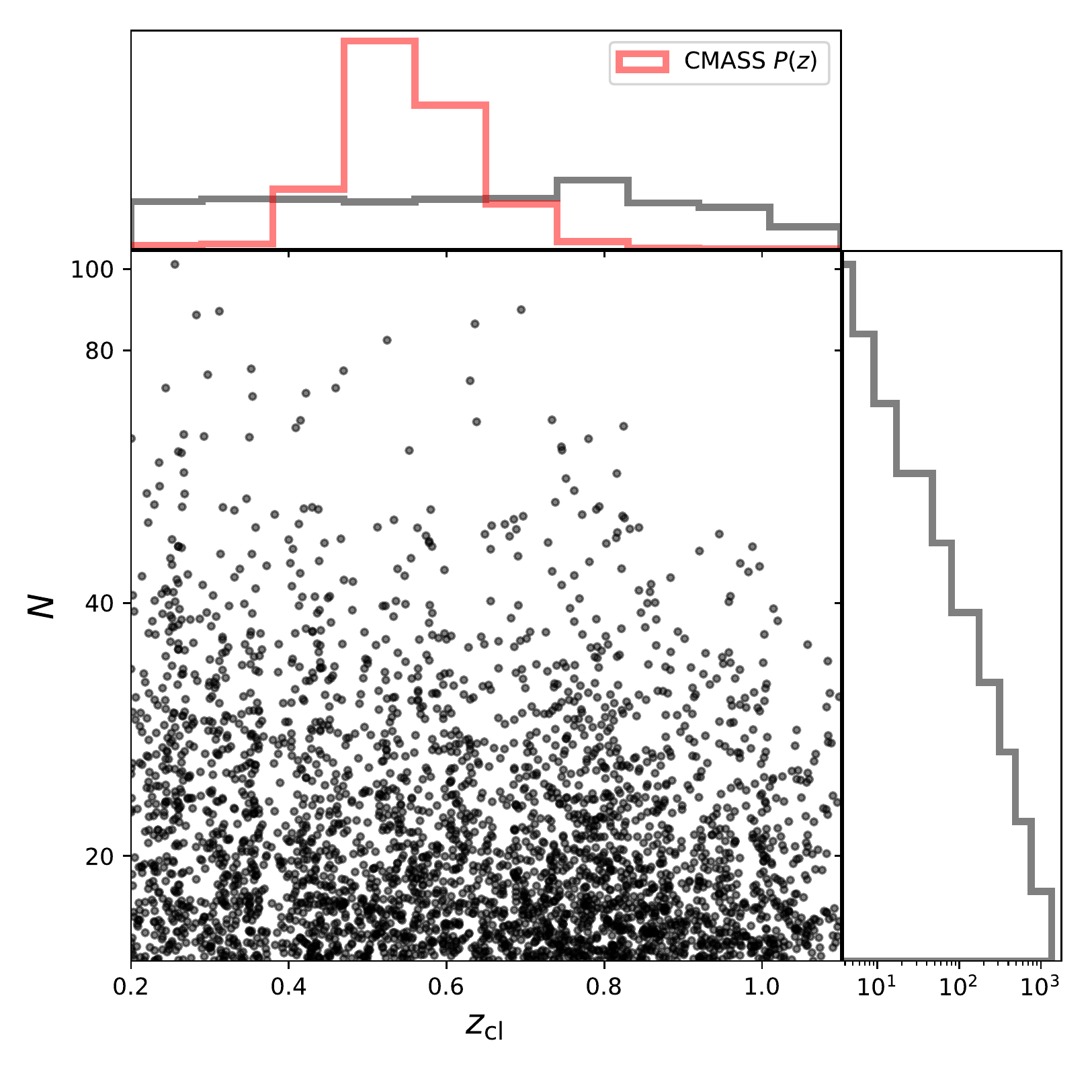}
}
\caption{
Distributions of richness \rich\ and redshift \zcl\ of the CAMIRA cluster sample used in this work.
CAMIRA clusters with $\rich\geq15$ at $0.2 \leq \redshift < 1.1$ are shown.
The upper (right) histogram shows the distribution of the cluster redshifts (the observed richness).
The normalized redshift distribution of CMASS galaxies are shown as the red curve in the upper subplot.
}
\label{fig:richz}
\end{figure}
\begin{figure*}
\centering
\resizebox{\textwidth}{!}{
\includegraphics[scale=1]{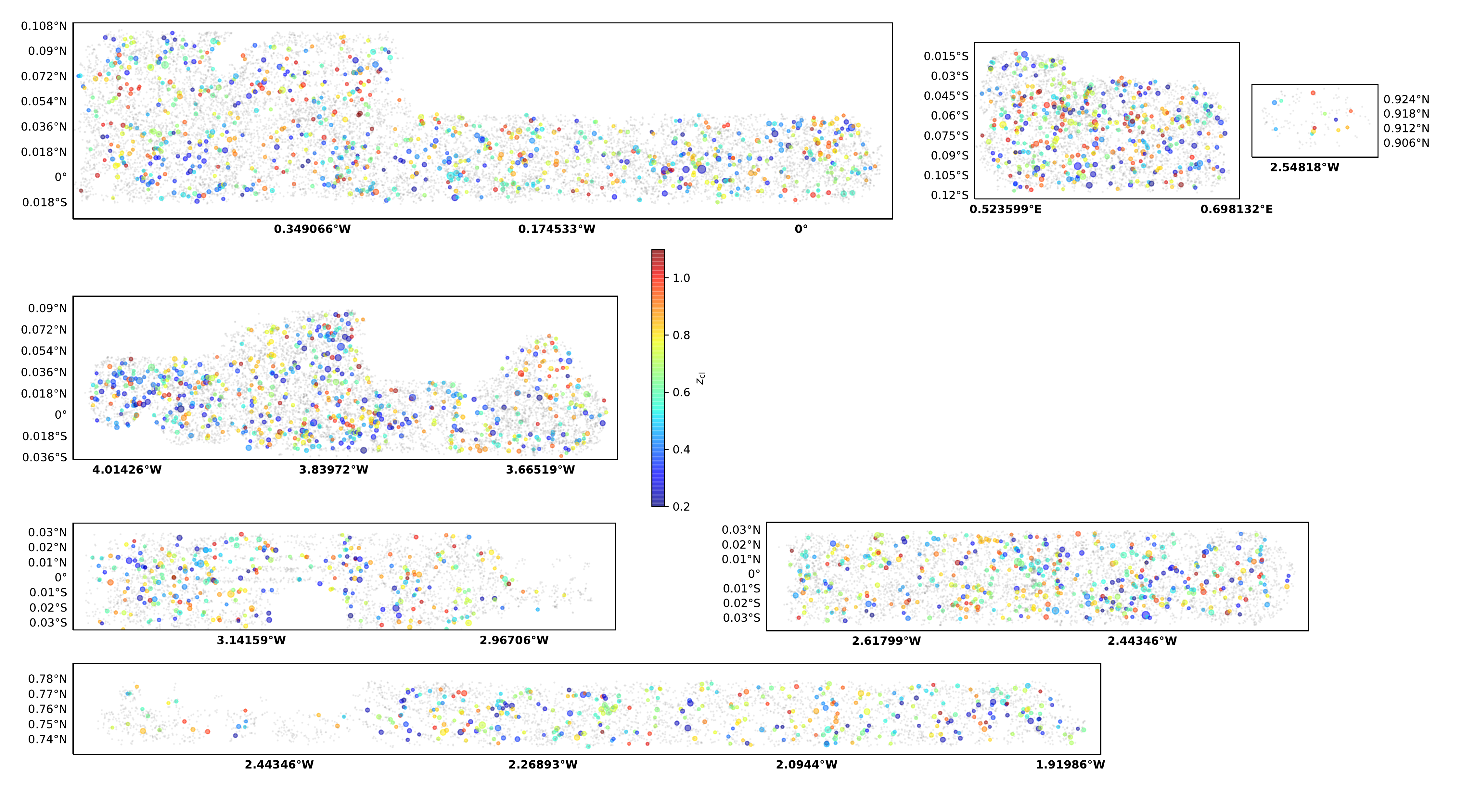}
}
\caption{
Angular distributions of CAMIRA clusters and CMASS galaxies in the common footprint between the HSC survey and the BOSS.
The CAMIRA clusters are shown as the circles color-coded by their redshifts \zcl\ with sizes proportional to their observed richness \rich.
The underlying grey points are the CMASS sample.
}
\label{fig:clustersample}
\end{figure*}

\section{Data}
\label{sec:data}

A brief overview of the HSC survey is given in Section~\ref{sec:hsc}.
In this work, we make use of 
the optically selected clusters
from the HSC Survey and the CMASS galaxy sample from the BOSS survey, as described in Section~\ref{sec:cluster_sample} and Section~\ref{sec:cmass}, respectively.
Meanwhile, we construct the mock catalogs for both samples of clusters and CMASS galaxies using a large set of N-body simulations, as detailed in Section~\ref{sec:mocks}.

\subsection{The HSC survey}
\label{sec:hsc}

The HSC survey is an imaging survey in the framework of a Subaru Strategic Program to image a sky area of 1400\,deg$^2$ in five broadband filters ($grizy$).
The imaging is carried out using the wide-field camera Hyper Suprime-Cam \citep{miyazaki15,miyazaki18} installed on the 8.2\,m Subaru Telescope. 
There are three layers in the HSC survey: WIDE, DEEP and UltraDEEP.
In the interest of a large and uniform coverage on the sky, we only use the data from the WIDE layer for constructing the cluster catalog (see Section~\ref{sec:cluster_sample}).
The imaging reduction and catalog construction are processed by the \texttt{hscPipe} \citep{bosch18}, for which the performance of photometric measurements is fully verified in \cite{huang18}.

In this work, we make use of the S18A data set from the HSC survey to construct the cluster catalog.
We have applied the bright star masks modified from \cite{coupon18} to the footprint, because a different scheme of background subtraction is used in cataloging the S18A data \citep[for more details, see][]{hscpdr2}.

\subsection{Cluster sample}
\label{sec:cluster_sample}

In this work, we make use of the cluster sample constructed by the \texttt{CAMIRA} algorithm \citep[Cluster finding Algorithm based on Multi-band Identification of Red-sequence gAlaxies;][]{oguri14} in the HSC Survey.
We refer interested readers to \cite{oguri14} and \cite{oguri18} for more details of the \texttt{CAMIRA} cluster finder.
In what follows, a brief overview of the \texttt{CAMIRA} algorithm is given.

\texttt{CAMIRA} is a matched-filter and red-sequence based cluster finder that relies on the stellar population synthesis model with the aid of calibration using spectroscopically confirmed galaxies.
After identifying a galaxy cluster, \texttt{CAMIRA} assigns a photometric redshift estimate and a richness \rich, which is an effective number of galaxy members used as the cluster mass proxy, to the system.
The center of each cluster is identified as the location of the Brightest Cluster Galaxy (BCG), which is suggested to be a good representative of the cluster center given the small offset ($\lesssim0.1$ Mpc/$h$ in the physical coordinate) between the BCGs and the X-ray peaks \citep{oguri18}.
Since the purpose of this work is to investigate the correlation functions in the comoving coordinate on a scale $\gtrsim10$ Mpc/$h$, this level of mis-centering is negligible.

In this work, we use the CAMIRA cluster catalog in the Full-Depth-Full-Color (FDFC) footprint of the HSC WIDE layer with area of $\approx427$~deg$^2$ (after applying the bright
star mask).
In this cluster catalog, we further select the clusters with richness $\rich\geq15$ at redshift $0.2\leq\redshift<1.1$ in the interest of consistency with previous work.
Specifically, the cluster sample selected in this criteria was previously studied using weak-lensing shearing \citep{murata19} and magnification \citep{chiu20}.
Therefore, this choice of the cluster selection enables a direct comparison with the results from gravitational lensing.
After the selection in richness and redshift, we further apply the mask of the CMASS galaxy sample (see Section~\ref{sec:cmass}) that we will cross-correlate with, such that the footprints of the cluster and CMASS samples are identical.
This reduces the area to $\approx403$~deg$^2$.
As a result, the final cluster sample consists of 3057 systems with $\rich\geq15$ at $0.2\leq\redshift<1.1$, which are shown as the 
points in Figure~\ref{fig:richz}.
Their distribution on the sky is shown as in Figure~\ref{fig:clustersample}.
Note that the redshift distribution of CAMIRA clusters is flatter than those of theoretical predictions.
This could be explained by a redshift-dependent scatter in richness at fixed mass, as proposed in \cite{murata19}, where they found that the scatter in richness has a quadratic behavior with the lowest value at the intermediate redshift ($\redshift\approx0.5$).
Higher scatter results in more clusters at given redshift, such that this redshift-dependent scatter produces a flat redshift distribution of clusters.
This redshift-dependent scatter was not statistically significant (at a level of $\lesssim2\sigma$), we therefore do not include this in our modelling of scatter in richness.
We leave a more complex modelling of the scatter to the future work.

Given the quality of the HSC data sets, the performance of the photometric redshift (\zcl) estimation of CAMIRA clusters is remarkable out to redshift $\redshift\lesssim1.2$.
Specifically, the bias and scatter in terms of $(\zcl - z_{\rm{BCG}})/(1 + z_{\rm{BCG}})$ are quantified to be $-0.0013$ and $0.0081$, respectively, where $z_{\rm{BCG}}$ is the observed spectroscopic redshift of the BCGs.
In the interest of uniformity, we use the photometric redshift \zcl\ for each cluster, regardless of available spectroscopic redshifts.
We note that a detailed modeling of photo-\redshift\ uncertainties is needed to correctly interpret observed correlation functions in redshift-space (see Section~\ref{sec:basics}), even with this sub-percent level of precision in the redshift estimation.

\subsubsection*{Cluster random catalogs}
\label{sec:random}

To calculate correlation functions from a given survey, we need to build random catalogs with the same survey geometry and 
redshift distribution as
the data.
To construct the random catalog for CAMIRA clusters, we follow a similar procedure as described in \citet{baxter16} to take into account the survey geometry, bright star masks, and the mask due to the CAMIRA cluster finder.
Specifically, for a given redshift we randomly draw a point in the survey footprint, excluding the regions indicated by the star mask, to investigate whether a CAMIRA cluster could be detected at this point.
If the cluster could be detected,  we add this point into the random catalog.
We repeat this process until the size of the random catalog reaches 40 times larger than the CAMIRA cluster catalog.
This procedure guarantees a random angular distribution  while taking into account the masks due to bright stars and the cluster finder.

We stress that the spatial filter of the CAMIRA algorithm is independent of richness \citep{oguri14}, so the mask of the CAMIRA finder only depends on the cluster redshift.
For the default analysis, we create the random catalog at redshift of $0.56$, which is approximately the mean redshift of the CAMIRA cluster sample.
We also create two cluster random catalogs at lower and higher redshifts ($0.26$ and $0.84$ respectively), and find that the change to the default analysis is negligible compared to the statistical uncertainty (see Appendix~\ref{sec:sys} for more details).
This is expected, because the spatial filter of the CAMIRA algorithm has only mild redshift dependence \citep{oguri14,oguri18}.

After randomizing the angular distribution of the random catalog, we then sample a random distribution along the line-of-sight direction based on the data.
Specifically, we bootstrap the observed redshifts of CAMIRA clusters (with replacements) and assign it to each point in the random catalog.
This is referred to as the ``shuffled'' redshift in Section~6 of \cite{ross12}.
This ensures that the redshift distribution of the random catalog includes not only the redshift uncertainty of observed clusters, but also the systematics introduced by the cluster finding algorithm \citep[e.g., the filter transition effect;][]{rykoff14,soergel16}.
As a result, the random catalog with a density of $1000$ points per square degree (or $\approx40$ times larger than the data) is obtained.

We find that sampling the redshift estimates to the random points following the 
redshift distribution of CAMIRA clusters 
after smoothing using a Gaussian kernel
results in negligible difference compared to the current statistical uncertainty.
We refer readers to Appendix~\ref{sec:sys} for more details.

\subsection{The CMASS sample}
\label{sec:cmass}

We use the spectroscopic sample of galaxies from the Baryon Oscillation
Spectroscopic Survey \citep[BOSS;][]{dawson13}, which is the largest survey in the Sloan Digital Sky Survey-III \citep[SDSS-III;][]{eisenstein11} program.
In the BOSS, there are two galaxy samples: LOWZ and CMASS.
The LOWZ sample targets the low-redshift galaxy population at $\redshift\lesssim0.4$, mainly dominated by Luminous Red Galaxies \citep{eisenstein01}.
On the other hand, the CMASS sample targets galaxies at high redshift of $0.4\lesssim\redshift\lesssim0.7$, which are pre-selected by using the imaging of the SDSS-II program \citep{aihara11}.
The target selection in the CMASS sample is based on a combination of customized color and magnitude cuts, such that a sample of galaxies with approximately ``constant stellar mass'' is expected.
We refer readers to \cite{rodriguez-torres16} for more details of the selection of CMASS galaxies.

In this work, we focus on the CMASS sample (in
both north and south Galactic Caps) from the Data Release 12 \citep[DR12;][]{alam15}, as the final data release\footnote{https://www.sdss.org/dr15/spectro/lss/\#BOSS} of the SDSS-III.
We only use the regions that are overlapping the S18A FDFC footprint of the HSC survey.
Moreover, we carefully apply the bright star mask of the HSC survey to the CMASS catalog, such that the footprints of the CMASS and cluster samples share the same geometry on the sky.
The common footprint between the CAMIRA and CMASS samples has area of $\approx403$~deg$^2$, in which about $37k$ CMASS galaxies are present with a median redshift of $\redshift_{\mathrm{BOSS}}=0.57$.
The normalized redshift distribution of CMASS galaxies is shown as the red curve in Figure~\ref{fig:richz}.
Their distribution on the sky is shown as in Figure~\ref{fig:clustersample}.

For the random catalog of observed CMASS galaxies, we make use of the random catalogs from both the north and south Galactic Caps that are publicly available\footnote{https://www.sdss.org/dr15/spectro/lss/\#BOSS} from the DR12.
These random catalogs already account for the angular mask of the BOSS.
For the random catalogs from both northern and southern hemispheres, we first select the common footprint between the BOSS and HSC survey and then exclude the regions indicated by the HSC masks due to bright stars.
Then, we randomly draw a redshift estimate from the observed CMASS sample (with replacements) and assign it to each point in the random catalog, separately for both hemispheres.
This is identical to the construction of the CMASS random catalog as in
\citet[][see also Section~\ref{sec:cluster_sample}]{ross12}.
In the end, the final random catalog for the observed CMASS galaxies is obtained by combining the random catalogs from both northern and southern hemispheres.
This process naturally accounts for the difference in the observed redshift distributions between the north and south Galactic Caps.

\subsection{Mock catalogs}
\label{sec:mocks}

In this work, we make use of the mock halo catalogs from the N-body simulations in \citet[][]{takahashi17} for the tasks of (1) the construction of covariance matrices of the correlation functions, and (2) the end-to-end validation of the codes.
We will detail these two tasks in Section~\ref{sec:covar} and Section~\ref{sec:modeling}, respectively.
In what follows, a brief summary of the mock catalogs is given.
We refer readers to \citet{takahashi17} for more details of the mock halo catalogs.

A number of 108 full-sky cosmological N-body simulations with high resolution is presented in \citet{takahashi17} under
a framework of the standard flat $\LCDM$ cosmology with $\Omega_{\mathrm{M}}=0.279$, $\Omega_{\Lambda}=0.721$, $\sigma_{8}=0.82$, and $H_0 = 70\mathrm{km/s}/\mathrm{Mpc}$, which is consistent with the \WMAP9 result \citep{hinshaw09}.
Each set of the N-body simulations is performed using the \texttt{GADGET2} code \citep{springel01,springel05} in the order of nested cubic boxes around an observer with different box sizes, ranging from $450h^{-1}\mathrm{Mpc}$ to $6300h^{-1}\mathrm{Mpc}$ with a step of $450h^{-1}\mathrm{Mpc}$. 
Each box contains $2048^3$ particles, which corresponds to a particle mass ranging from $\approx10^{9}h^{-1}\Msun$ to $\approx10^{12}h^{-1}\Msun$, depending on the box side.
The resulting matter power spectra of the N-body simulations are fully resolved on the scale of $k<5h\mathrm{Mpc}^{-1}$ at $\redshift<1$ and are in good agreement with the theoretical model predicted by the \texttt{Halofit} \citep{smith03, takahashi12}.
The dark matter halos are identified by the \texttt{ROCKSTAR} halo finder \citep{behroozi13} with a criterion that the minimal halo mass must be at least 50 times of the particle mass.
In this configuration, the resulting mock halo catalogs effectively cover the halo mass range of CMASS galaxies and CAMIRA clusters, by design.
In the mass and redshift range of interested in this work, the mass function and the linear halo bias of mock halos are verified to agree with those predicted by the \cite{tinker08} and \cite{tinker10} formulas within $12\percent$ and $10\percent$, respectively.
In \citet[][see also their Figure~21]{takahashi17}, they quantified that the systematic difference in the linear halo bias between the simulated halos and those predicted by the \cite{tinker10} fitting formula is at a level of $\lesssim10\percent$ at the wavenumber $k<0.4~h/\mathrm{Mpc}$ ($k<0.1~h/\mathrm{Mpc}$) for $M_{200\mathrm{m}} = 10^{13}\Msun$ ($M_{200\mathrm{m}} = 10^{14}\Msun$) at $\redshift<1.4$ ($\redshift<1.5$).
In other words, we expect a systematic uncertainty in the linear halo bias of mock halos at a level of $10\percent$ in this work.

With proper rotations of the sky, from each mock catalog we further 
tile
four non-overlapping footprints with the same geometry of the common footprint between the BOSS and HSC survey.
This ensures that the four catalogs from the same full-sky simulation are nearly independent with each other on the scale smaller than 
the current footprint of the survey studied in this work.
As a result, we make use of a total number of 432$(=108\times4)$ mock catalogs that are able to represent the realistic properties, e.g., the mass function and halo clustering, of CMASS galaxies and CAMIRA clusters.
From each mock halo catalog among these 432, we further construct a mock catalog of CAMIRA clusters and a mock catalog of CMASS galaxies, as detailed in Section~\ref{sec:mockcluster} and Section~\ref{sec:mockcmass}, respectively.

\subsubsection{Mock clusters}
\label{sec:mockcluster}

To utilize these mock catalogs in the same way as we do for CAMIRA clusters, we need to assign the mass proxy, i.e., richness \rich, for each mock halo.
Specifically, we assign a richness estimate to each mock halo following the \rtm\ relation defined in equation~(\ref{eq:richness_to_mass}) with the parameters constrained by gravitational lensing.
We use the parameter obtained in \cite{chiu20}, in which they constrained the \rtm\ relation of CAMIRA clusters in the same richness and redshift ranges using weak-lensing magnification (flux magnification bias) alone.
To be exact, for each mock catalog we sample a set of the \rtm\ parameters from the chain of the lensing magnification constraint\footnote{The ``Joint'' constraints in the fifth column of Table~1 in \cite{chiu20}}.
Then, we assign a richness estimate to each mock halo in this catalog accounting for the intrinsic scatter and measurement uncertainty of richness given the true halo mass.
We repeat this procedure for 432 mock halo catalogs.
The mean values of the \rtm\ relation used in the richness assignment are
\[
(\Arich, \Brich, \Crich, \Drich) = (17.72, 0.92, -0.48, 0.15) \, .
\]
Note that these parameters are defined in equation~(\ref{eq:richness_to_mass}).
We stress that the \rtm\ relation from lensing magnification is statistically consistent ($\lesssim1.8\sigma$) with the result independently obtained by combining weak shear and cluster abundance \citep{murata19}.
By marginalizing over the parameter constraint from lensing magnification, we effectively takes into account the uncertainty of the \rtm\ relation in the richness assignment.

After assigning the estimate of richness, we further perturb the redshift $z_{\rm{mock}}$ with respect to the true redshift $z_{\rm{true}}$ of mock halos.
This is done by following a Gaussian distribution with a zero mean and the scatter with a form of $\sigma_{(z_{\rm{mock}} - z_{\rm{true}})/(1 + z_{\rm{true}})} = 0.0093\times\left(\rich/{\rich}_{\mathrm{piv}}\right)^{-0.18}$, depending on the cluster richness.
Note that this richness-dependent scatter in the redshift estimate is directly constrained by CAMIRA clusters (for more details, see Section~\ref{sec:model_photoz}), and that we specifically apply this to mock catalogs in order to mimic the observed photo-\redshift\ uncertainty.

Finally, we apply the richness and redshift cuts ($\rich\geq15$ and $0.2\leq\redshift<1.1$) to mock catalogs, as we do in the analysis of CAMIRA clusters.
This leads to 432 mock cluster catalogs.

We construct a random catalog for 432 mock cluster catalogs in a similar way as done in constructing the random catalog for CAMIRA clusters.
Specifically, we follow the same randomization of the angular distribution to account for various masks in the survey footprint, and then assign a redshift to each random point by bootstrapping the redshifts from the joint catalog of 432 mock cluster samples. 
That is, the only difference to the random catalog of observed CAMIRA clusters is that the random redshift estimate is ``shuffled'' from mock clusters.
It is important to note that we do not bootstrap the redshift from observed CAMIRA clusters, because (1) the mock clusters 
are constructed by using a different cluster finder algorithm rather than that based on cluster red-sequence (see Section~\ref{sec:cluster_sample}), and  (2) the redshift estimates of mock clusters only take into account measurement uncertainties but not the systematics that is subject to the CAMIRA cluster finder.
Meanwhile, we ensure that the resulting random catalog is at least $40$ times larger than each mock cluster catalog.

\subsubsection{Mock CMASS galaxies}
\label{sec:mockcmass}

In this study, we also construct a mock catalog of CMASS galaxies from each out of the 432 realizations.
First, we assume that the observed redshift $z_{\rm{mock}}$ of  mock halos is identical to the true redshift $z_{\rm{true}}$, ignoring measurement uncertainties.
This is a reasonable assumption, because the redshifts of CMASS galaxies are secured by spectroscopic data with negligible uncertainties.
Note that the redshift $z_{\rm{mock}}$ still includes the contribution from the peculiar velocity of halos.
That is, the RSD and FoG effects are present in the mock catalogs.

Second, we randomly select mock halos following the observed redshift distribution (denoted as $P_{\mathrm{CMASS}}\left(\redshift\right)$) of CMASS galaxies.
In practice, this is done by sampling a random variable uniformly distributed between 0 and $\mathtt{max}\left\lbrace P_{\mathrm{CMASS}}\left(\redshift\right)\right\rbrace$ to each mock halo, for which it is selected if the random variable is smaller than $P_{\mathrm{CMASS}}\left(\redshift_{\rm{mock}}\right)$ at the mock halo redshift $\redshift_{\rm{mock}}$.
We derive $P_{\mathrm{CMASS}}\left(\redshift\right)$ using an interpolation over the observed redshift distribution in a redshift interval of $\left[0,1.1\right]$ with a step of $0.001$.
This ensures that the selected mock CMASS galaxies have the same redshift distribution as the data.
Despite different redshift distributions of observed CMASS galaxies between the north and south Galactic Caps \citep{ross12},
we note that $P_{\mathrm{CMASS}}\left(\redshift\right)$ is derived based on the joint sample of CMASS galaxies from both hemispheres in the common footprint of the BOSS and the HSC survey.
That is, $P_{\mathrm{CMASS}}\left(\redshift\right)$ represents the effective redshift distribution of the CMASS sample in this work, as a whole. 

Third, we assign a probability of hosting a central galaxy to each mock halo by applying the prescription of halo occupation distribution (HOD) in \cite{manera13}, which is specifically designed for creating CMASS-like mock galaxies.
Specifically, the mean number of central galaxies for each mock halo reads
\[
\left\langle N_{\mathrm{cen}}\right\rangle = \frac{1}{2}\left[1 + \erf\left(\frac{\log M_{200\mathrm{m}} - 13.09}{0.596}\right)\right] \, .
\]
Then, each halo is selected with a probability equal to $\left\langle N_{\mathrm{cen}}\right\rangle$.
For example, a halo with $\left\langle N_{\mathrm{cen}}\right\rangle=0.5$ has a probability of $50\percent$ to be selected into the mock CMASS sample.
This selection using the HOD modeling effectively leads to a mock galaxy sample with a mass distribution that is consistent with that of observed CMASS galaxies, by design.

Last, only the halos satisfying the selection criteria of redshift and HOD modeling are selected into the final mock CMASS samples.
This procedure is repeated for the 432 mock realizations, resulting in the same number of mock CMASS catalogs.
Note that we assume no correlation between the mass and redshift in selecting mock CMASS galaxies in this approach.
The resulting mock CMASS galaxies have a mass range\footnote{Alternatively, this corresponds to $12.2\lesssim\log\left(\frac{M_{200\mathrm{m}}}{h^{-1}\Msun}\right)\lesssim14.2$ with a median value of $\approx13.0$} of $11.9\lesssim\log\left(\frac{\Mfiveoo}{h^{-1}\Msun}\right)\lesssim13.9$ with a median value of $\approx12.8$.
Using the formula in \cite{tinker10}, this implies that the halo bias has a range between $\approx1.1$ and $\approx4.2$ with a mean (median) value of $\approx1.9$ ($\approx1.7$), which is in good agreement with the observational result of CMASS galaxies \citep[$\approx1.93\pm0.17$;][]{chuang13}.
Given the systematic uncertainty at a level of $10\percent$ in the halo bias of mock halos (see Section~\ref{sec:mocks}), this implies good consistency between the resulting mock and observed CMASS galaxies.
Additionally, we also extract the information of the line-of-sight velocity dispersion $\sigma_{\mathrm{v}}$ from these mock CMASS catalogs; it is estimated as $\sigma_{\mathrm{v}}\approx310~$km/sec in the physical space.

We cannot directly use the random catalog constructed for observed CMASS galaxies in calculating the correlation functions of mock CMASS galaxies, because (1) the redshift distributions of observed CMASS galaxies are different between the north and south Galactic Caps \citep{ross12}, and (2) we select the mock galaxies following the effective redshift distribution $P_{\mathrm{CMASS}}(\redshift)$ of observed CMASS sample across the footprint (see Section~\ref{sec:mockcmass}).
That is, the random catalog of observed CMASS galaxies contains a redshift distribution dependent on the northern and southern caps,
which is not the case
for mock CMASS galaxies. 
Therefore, for each mock CMASS catalog, we build a random catalog by first sampling random points on the sky with the identical footprint of the data (after applying the star mask), then followed by the redshift assignment to each random point according to the effective redshift distribution $P_{\mathrm{CMASS}}(\redshift)$.
We ensure that the size of the resulting random catalog is at $\approx30$ times larger than each mock CMASS catalog.

%
%

\section{Measurements}
\label{sec:measurements}

In this work, we determine three correlation functions in the redshift space: auto-correlations of CAMIRA clusters and CMASS galaxies, and their cross-correlations.
In what follows, we detail our procedure to obtain the measurements, which will be used to calibrate the \rtm\ relation of CAMIRA clusters in a self-calibration manner, i.e., solely based on halo clustering, in Section~\ref{sec:modeling}.

\subsection{Correlation Functions}
\label{sec:measurements_of_correlations}

The two-point auto-correlation function of a tracer $X$, $\xi_{\mathrm{XX}}(s)$ is derived using the \citet{landy93} estimator, namely
\begin{equation}
\label{eq:landyszalayestimator_auto_3d}
\xi_{\mathrm{XX}}(s) = \frac{
\mathrm{D}\mathrm{D} - 2 \mathrm{D}\mathrm{R} + \mathrm{R}\mathrm{R}
}{
\mathrm{R}\mathrm{R}
} \, , 
\end{equation}
where $\mathrm{X}=\{\mathrm{c},\mathrm{g}\}$ and $\mathrm{c}$ and $\mathrm{g}$ stand for clusters and galaxies, respectively. 
Here,
$\mathrm{D}\mathrm{D}=\mathrm{D}\mathrm{D}(s)$,
$\mathrm{D}\mathrm{R} = \mathrm{D}\mathrm{R}(s)$, and
$\mathrm{R}\mathrm{R} = \mathrm{R}\mathrm{R}(s)$ are the normalized numbers of pairs with a separation of the redshift-space distance $s$ between the data-data, data-random, and random-random catalogs, respectively.
This estimator given by equation~(\ref{eq:landyszalayestimator_auto_3d}) can be generalized for the cross-correlation $\xi_{\mathrm{XY}}(s)$ as 
\begin{equation}
\label{eq:landyszalayestimator_cross_3d}
\xi_{\mathrm{XY}}(s) = \frac{
\mathrm{D_X}\mathrm{D_Y} - \mathrm{D_X}\mathrm{R_Y} - \mathrm{R_X}\mathrm{D_Y} + \mathrm{R_X}\mathrm{R_Y}
}{
\mathrm{R_X}\mathrm{R_Y}
} \, ,
\end{equation}
where
$\mathrm{D_X}\mathrm{D_Y}$,
$\mathrm{D_X}\mathrm{R_Y}$,
$\mathrm{R_X}\mathrm{D_Y}$, and
$\mathrm{R_X}\mathrm{R_Y}$ are, respectively, the normalized numbers of data-data, data-random, random-data, and random-random pairs
found at a separation of $s$ for tracers $\mathrm{X}$ and $\mathrm{Y}$.
All tasks of pair counting in this work is done by using \texttt{TreeCorr} \citep{jarvis04}.

Using equation~(\ref{eq:landyszalayestimator_auto_3d}) and (\ref{eq:landyszalayestimator_cross_3d}), we derive the following correlation functions in this work:
\begin{enumerate}
\item \xicamira: the auto-correlation of CAMIRA clusters,
\item \xicmass: the auto-correlation of CMASS galaxies, and
\item \xicamiracmass: the cross-correlation between CAMIRA clusters and CMASS galaxies.
\end{enumerate}
The correlation functions of \xicamira, \xicamiracmass\ and \xicmass\ are estimated in the redshift-space distance $s$ between $10~h^{-1}\mathrm{Mpc}$ and $60~h^{-1}\mathrm{Mpc}$ with logarithmic binning of 7 steps.

Except treating the CAMIRA clusters as a whole in calculating (i) and (iii), we also perform a ``subsample'' analysis to further investigate the clustering properties as functions of richness and redshift.
Namely, we split the cluster sample into 
two redshift bins ($0.2\leq\redshift<0.7$ and $0.7\leq\redshift<1.1$) and two richness bins ($15\leq\rich<25$ and $\rich\geq 25$), with four subsamples in total.
Then, we re-measure the clustering functions (i) and (iii) independently for each subsample.

\begin{figure*}
\centering
\resizebox{\textwidth}{!}{
\includegraphics[scale=1]{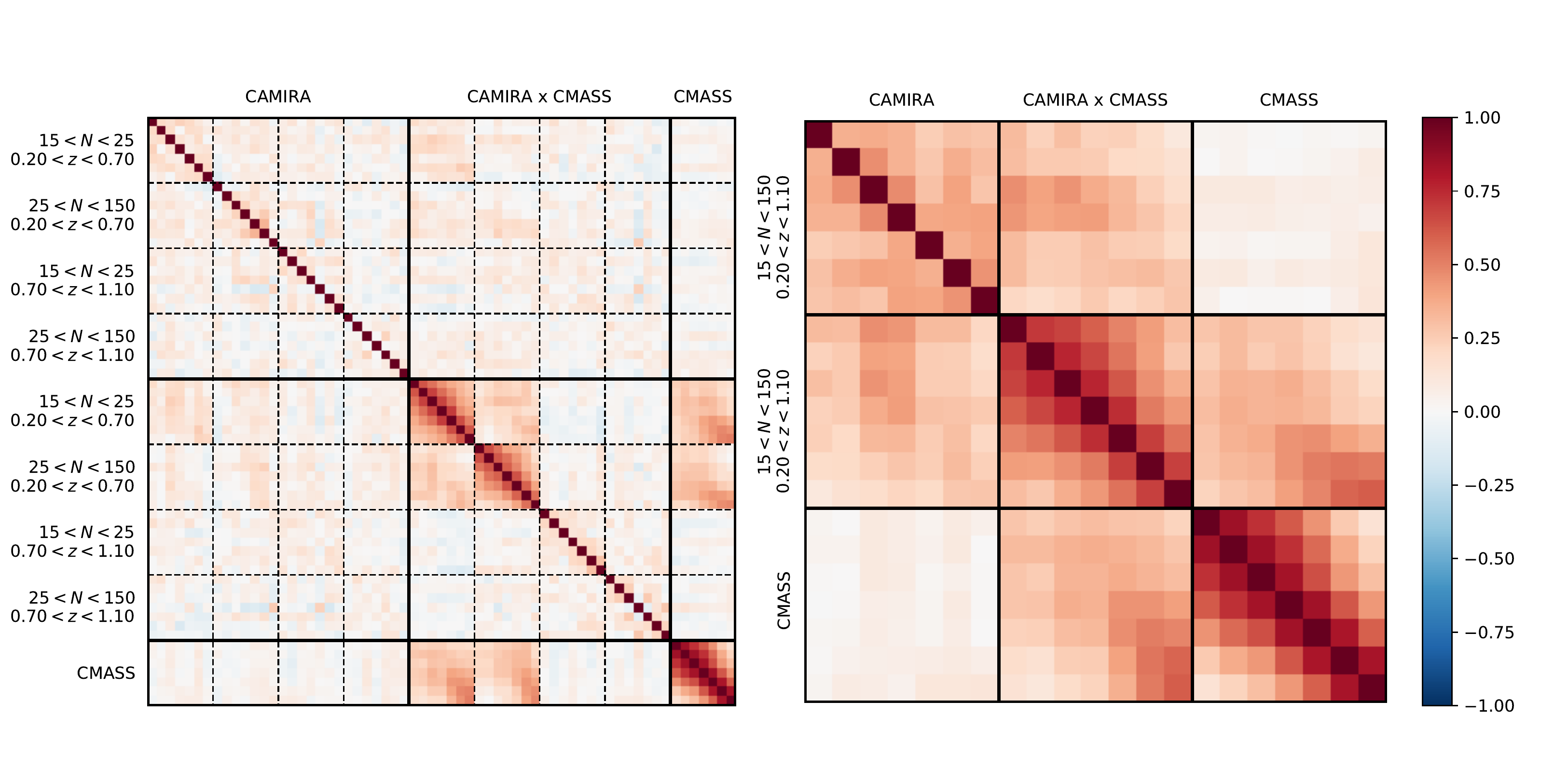}
}
\caption{
Left: Normalized covariance matrices (i.e, the correlation matrices) of the correlation functions from the subsamples, where the boxes enclosed by the dashed (solid) lines indicate the subsamples of different richness and redshift bins (the measurements of CAMIRA auto-correlation, $\mathrm{CAMIRA}\times\mathrm{CMASS}$ cross-correlation, and CMASS auto-correlation functions).
Right: Same as the left panel but measured from the whole cluster sample, where the boxes enclosed by the solid lines indicate the CAMIRA, $\mathrm{CAMIRA}\times\mathrm{CMASS}$, and the CMASS correlation functions.
The color-bar shows the strength of correlation coefficients.
}
\label{fig:corre_3d}
\end{figure*}

\subsection{Construction of covariance matrices}
\label{sec:covar}

Covariance matrices are needed for statistical analyses because different bins of the observed correlation functions are strongly correlated.
By taking the advantage of the large sizes
of N-body simulations, we derive covariance matrices based on the 432 mock halo catalogs, which are described in Section~\ref{sec:mocks}.

We first repeat the measurements (\xicamira, \xicamiracmass, and \xicmass) on each mock catalogs.
In this way, we generate 432 sets of the measurements in the identical configuration as the data.
Then, for any combination of a data vector, denoted as $\mathfrak{D}$, we can find the corresponding measurements ${\mathfrak{D}}_{(i)}$ from the $i$-th mock catalog and derive the covariance matrix as
\begin{equation}
\label{eq:covar_mock}
\mathfrak{C} = \frac{1}{N_{\mathrm{mock}} - 1}
\sum_{i=1}^{N_{\mathrm{mock}}}
\left( {\mathfrak{D}}_{(i)} - \mathfrak{D}_{(\cdot)} \right)^{\mathrm{T}} \cdot
\left( {\mathfrak{D}}_{(i)} - \mathfrak{D}_{(\cdot)} \right) \, ,
\end{equation}
where $\mathfrak{D}_{(\cdot)} = \frac{1}{N_{\mathrm{mock}}}\sum_{i=1}^{N_{\mathrm{mock}}} {\mathfrak{D}}_{(i)}$, and
$N_{\mathrm{mock}} = 432$.
The resulting covariance matrix normalized by the diagonal elements, $\mathfrak{C}_{ij} / (\mathfrak{C}_{ii}\cdot \mathfrak{C}_{jj})^{1/2}$, is shown in Figure~\ref{fig:corre_3d}.
We found that the diagonal term of covariance matrices are stable after the number of realizations exceeds $\approx150$, suggesting that our covariance matrices are converged at the current amount of realizations (i.e, 432).

We further multiply a factor of 
$\frac{N_{\mathrm{mock}} - N_{\mathrm{D}} - 2}{N_{\mathrm{mock}} - 1}$, 
where $N_{\mathrm{D}}$ is the length of a data vector, to the inverse covariance matrix $\mathfrak{C}^{-1}$
to account for the underestimation of the uncertainty because of a finite number of realizations used in estimating the covariance matrix \citep{hartlap07}.
In this work, the correction factor, $\frac{N_{\mathrm{mock}} - 1}{N_{\mathrm{mock}} - N_{\mathrm{D}} - 2}$, ranges from $\approx1.9\percent$ (for the modeling of a correlation function of the whole sample; $N_{\mathrm{D}} = 7$) to $\approx17\percent$ (for a joint modeling of the auto- and cross-correlation functions in the ``subsample'' analysis; $N_{\mathrm{D}} =7 \times 4 ~\mathrm{subsamples} \times \left( 1 ~\mathrm{CAMIRA~auto} +1 ~\mathrm{cross} \right) + 7 \times \left(1~\mathrm{CMASS~auto}\right)= 63$).

%
%

\section{Modeling}
\label{sec:modeling}

We use the approach as described in \cite{sereno15} to model the correlation functions of \xicamira, \xicamiracmass, and \xicmass.

In this paper, we consider only the angle-averaged monopole component of a correlation function, 
\begin{equation}
\label{eq:monopole}
\xi_{\mathrm{XY}}(s) = \int{ \frac{k^2\dif k}{2\pi^2} j_{0}(ks) P_{\mathrm{XY}}(k)} \, ,
\end{equation}
in which $\mathrm{XY} = \{\mathrm{cc},\mathrm{cg},\mathrm{gg}\}$, $j_{0}$ is a zero-order spherical Bessel function, and $P_{\mathrm{XY}}(k)$ is the angle-averaged power spectrum.

In the presence of only the redshift-smearing (or FoG) and Kaiser effects, a redshift-space correlation function depends on the two directions that are  perpendicular to and along the line of sight, respectively.
Therefore, it is common to re-express the power spectrum in a polar coordinate of $\mathbf{k}=(k, \mu)$, where $\mu$ is the cosine angle of the vector $\mathbf{k}$ with respect to the line of sight.
In this way, the term $P_{\mathrm{XY}}$ has a generic form \citep[e.g.,][]{park94,peacock94,okumura15}, 
\begin{align}
\label{eq:generalform}
P_{\mathrm{XY}}
\left(
\mathbf{k}
\right) 
=
P_{\mathrm{m}}(k)
&\times
\left(
b_{\rm X} + \growthrte(\redshift_{\mathrm{X}})\mu^2
\right)
\left(
b_{\rm Y} + \growthrte(\redshift_{\mathrm{Y}})\mu^2
\right) \nonumber \\
&\times
G_v(k\mu\sigma_{v\mathrm{LoS},\mathrm{X}})
G_v(k\mu\sigma_{v\mathrm{LoS},\mathrm{Y}})
\nonumber \\
&\times
G_z(k\mu\sigma_{z\mathrm{LoS},\mathrm{X}})
G_z(k\mu\sigma_{z\mathrm{LoS},\mathrm{Y}})
\, ,
\end{align}
where
$P_{\mathrm{m}}(k)$ is the matter power spectrum;
the term $\left(\biash + \growthrte(\redshift)\mu^2\right)$ contains the \citet{kaiser1987} term describing the linear RSD effect;
$b_{\rm X}$ is the linear bias of halos which host a tracer;
$\growthrte(\redshift_{\mathrm{X}})$ is the growth rate evaluated at the redshift $\redshift_{\mathrm{X}}$ of a tracer $\mathrm{X}$;
the functions $G_v$ and $G_z$ are the damping functions caused by the nonlinear velocity dispersion ($\sigma_{v\mathrm{LoS}}$)
and photo-\redshift\ uncertainties ($\sigma_{\redshift\mathrm{LoS}}$), respectively. 
We describe each term as follows.

\begin{figure}
\centering
\resizebox{0.45\textwidth}{!}{
\includegraphics[scale=1]{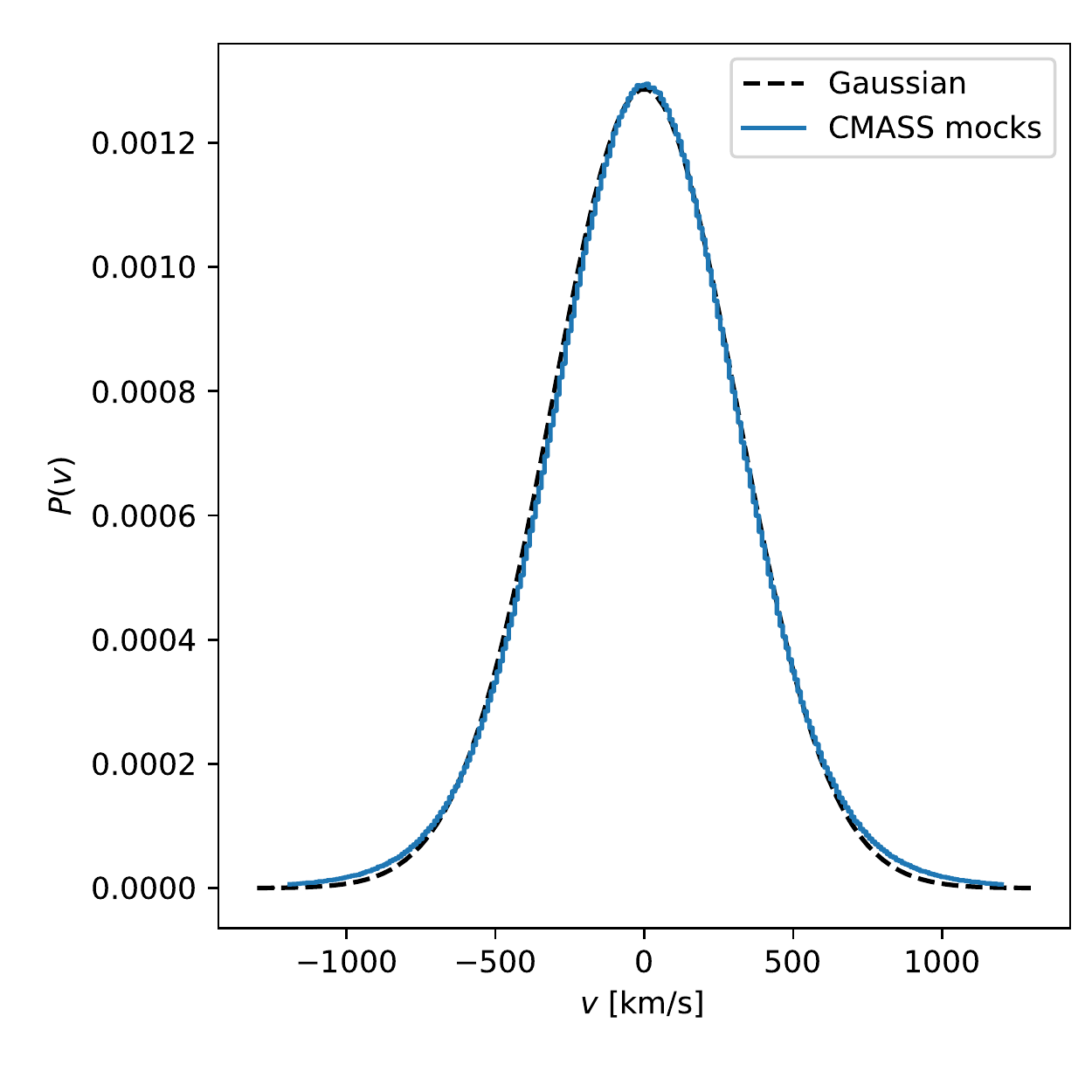}
}
\caption{
Distribution of the line-of-sight peculiar velocity $v$.
The normalized distribution of $v$, estimated based on the stacked catalog of 432 mock CMASS samples, is shown as the blue histogram.
This can be well described by a Gaussian distribution (dashed line) with a zero mean and the best-fit dispersion $\sigma_{v}=310~\mathrm{km}/\mathrm{sec}$.
This value is used in modeling the effect arising from the peculiar velocity of CMASS galaxies (see Section~\ref{sec:modeling_fog}).
}
\label{fig:vp}
\end{figure}

\subsection{Modeling of the Finger-of-God effect}
\label{sec:modeling_fog}

In this study, we assume a Gaussian function for the nonlinear smearing effect due to the line-of-sight velocity dispersion, $\sigma_{v}$, as
\begin{equation} 
\label{eq:model_fog}
G_v(k\mu\sigma_{v\mathrm{LoS}}) = 
\exp
\left(
\frac{-k^2\mu^2\sigma_{v\mathrm{LoS}}^2
}{2}
\right)
\, ,
\end{equation} 
where
\begin{equation} 
\label{eq:model_fog_2}
\sigma_{v\mathrm{LoS}} = 
\frac{
\sigma_{v} \left(1 + \redshift\right)
}{
H\left(\redshift\right)
}
\, .
\end{equation}
This model is supported by our mocks, in which the distribution of peculiar velocity along the line of sight indeed can be well described as a Gaussian distribution.
Figure~\ref{fig:vp} shows the distribution of the line-of-sight peculiar velocity of mock CMASS galaxies, which is well characterized by a Gaussian distribution with the best-fit dispersion of $\sigma_{v}=310$~km/s. 
We have carefully checked that the final result is not affected by the choice of the functional form of $G_v$ because we analyze the clustering data only on large scales, $s>10$~Mpc$/h$. 
For clusters for which only photo-\redshift\ data are available, not only the nonlinear velocity dispersion is smaller than that for galaxies but also the effect of photo-\redshift\ uncertainties is much severer.
Thus, this term can be ignored for our CAMIRA cluster sample (see Section~\ref{sec:basics}).

\subsection{Modeling of Photo-\redshift\ Uncertainties}
\label{sec:model_photoz}

We model the dispersion $\sigma_{\Delta\redshift}$ of photo-\redshift\ uncertainties in equation~(\ref{eq:generalform}) by a Gaussian distribution, described as
\begin{equation} 
\label{eq:model_photoz}
\sigma_{z\mathrm{LoS}}(k\mu\sigma_{z\mathrm{LoS}}) = 
\exp\left(
\frac{
-k^2
\mu^2
\sigma_{\redshift\mathrm{LoS}}^2
}{2}
\right)
\, ,
\end{equation} 
where
\begin{equation} 
\label{eq:model_photoz_2}
\sigma_{\redshift\mathrm{LoS}} = 
\frac{c
}{
H\left(\redshift\right)
}
\times
\sigma_{\Delta\redshift} 
\, ,
\end{equation}
following equation~(\ref{eq:sigmascale}).
This is a reasonable assumption, given that the measurement uncertainty of photo-\redshift\ is indeed distributed as a Gaussian in our work.
Thus, $G_v$ and $G_z$ eventually have the same functional form.

\begin{figure*}
\centering
\resizebox{0.48\textwidth}{!}{
\includegraphics[scale=1]{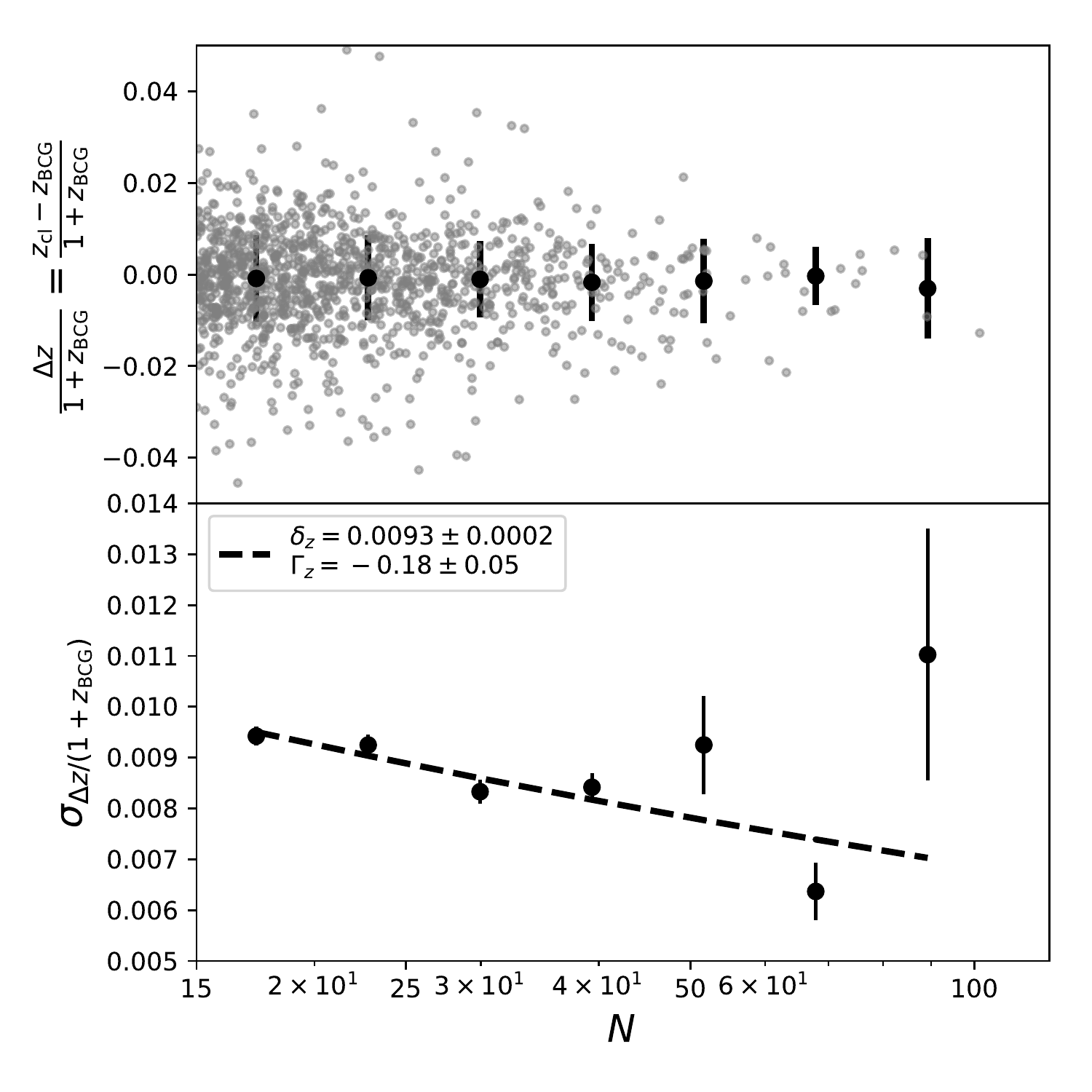}
}
\resizebox{0.48\textwidth}{!}{
\includegraphics[scale=1]{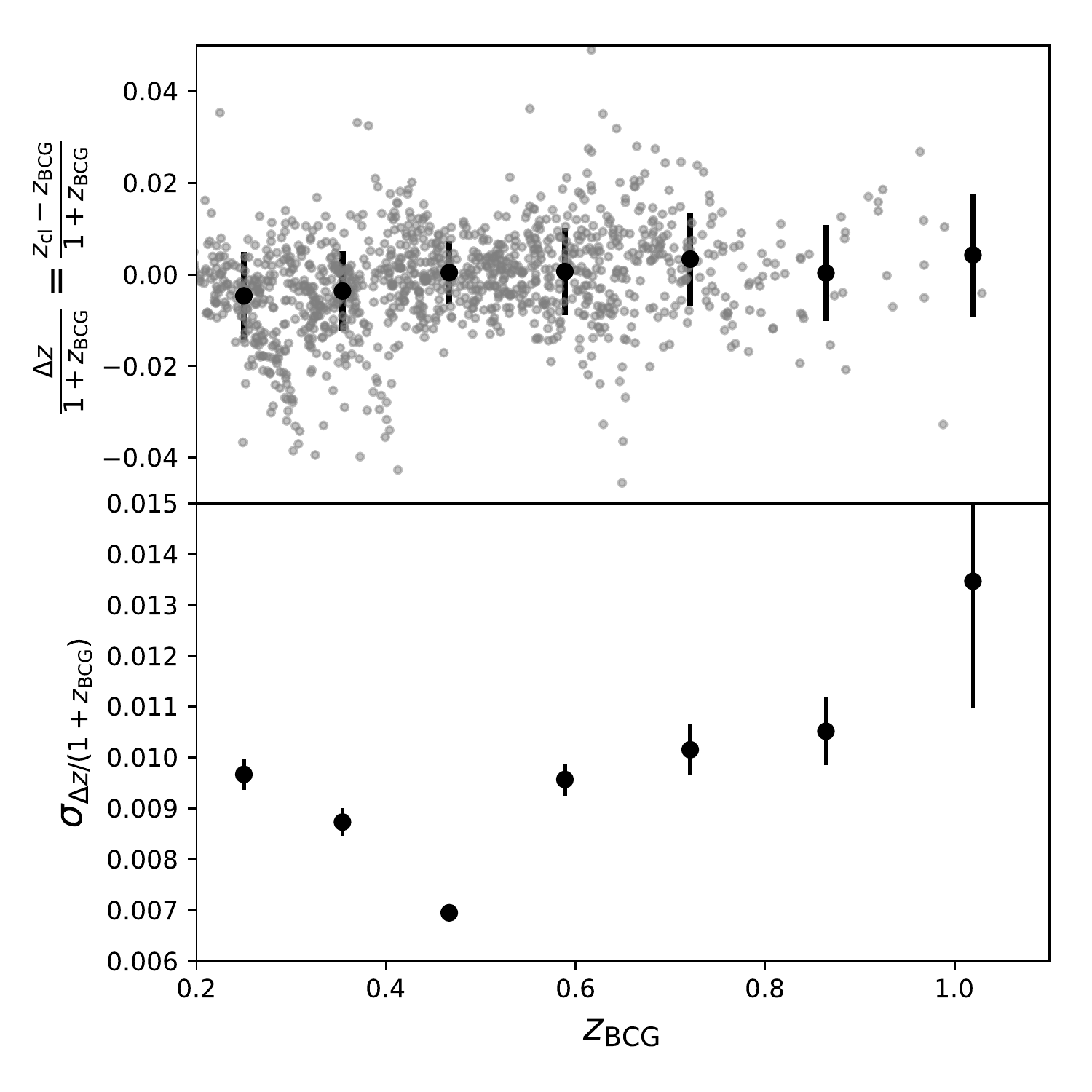}
}
\caption{
Distribution and the dispersion of the photo-\redshift\ uncertainty, which is characterized in terms of $\Delta~\redshift/\left(1 + {\redshift}_{\mathrm{BCG}}\right)$ with $\Delta~\redshift\equiv\zcl\ - {\redshift}_{\mathrm{BCG}}$, as a function of the cluster richness (left) and redshift (right).
In the left panel, the distribution of the photo-\redshift\ uncertainty is shown as the grey points in the upper plot.
The black circles represent the distributions of the photo-\redshift\ uncertainty in seven richness bins, assuming Gaussian distributions, with their errorbars as the size of dispersion.
The best-fit values and uncertainties of the Gaussian dispersion $\sigma_{\Delta~\redshift/\left(1 + {\redshift}_{\mathrm{BCG}}\right)}$ in seven richness bins are shown in the lower plot, for which we model them as a function of cluster richness by a power-law function (the dashed line).
In the right panel, we show the distribution of the photo-\redshift\ uncertainty (upper) and the sizes of their Gaussian dispersion (lower) as functions of cluster redshift, following the same configuration as in the left panel. 
We observe a non-monotonic behavior of $\sigma_{\Delta~\redshift/\left(1 + {\redshift}_{\mathrm{BCG}}\right)}$ in cluster redshift (see more discussions in Section~\ref{sec:model_photoz}).
}
\label{fig:photoz}
\end{figure*}

We derive the dispersion $\sigma_{\Delta\redshift}$ of photo-\redshift\ uncertainties $\Delta\redshift$ for CAMIRA clusters as follows. 
We assume that cosmological redshift, $\redshift_{\mathrm{c}}$, is identical to the redshift of the BCG, $\redshift_{\mathrm{BCG}}$, ignoring peculiar velocity as a subdominant factor. 
That is,  $\Delta\redshift\equiv\zcl - \redshift_{\mathrm{c}} = \zcl - \redshift_{\mathrm{BCG}}$.
Next, we model $\sigma_{\Delta\redshift/\left(1+\redshift_{\mathrm{BCG}}\right)}$ using a power law of cluster richness:
\begin{equation}
\label{eq:zdisp}
\sigma_{\Delta\redshift/\left(1+\redshift_{\mathrm{BCG}}\right)}(N) \equiv \frac{\sigma_{\Delta\redshift}}{1+\redshift_{\mathrm{BCG}}} = \delta_{\redshift}\left(\frac{\rich}{N_{\mathrm{piv}}}\right)^{\Gamma_{\redshift}} \, .
\end{equation}
with two free parameters $(\delta_{\redshift}, \Gamma_{\redshift})$ that can be constrained from the data.
Specifically, we bin CAMIRA clusters in seven richness bins, in which
the photo-\redshift\ uncertainty in terms of $\Delta\redshift/(1 + z_{\rm{BCG}})$ in the richness bin ${\rich}_{i}$ is modeled by a Gaussian distribution with the dispersion of $\sigma_{\Delta\redshift/\left(1+\redshift_{\mathrm{BCG}}\right)}({\rich}_{i})$ where $i=1,\cdots, 7$.
Then, we fit equation~(\ref{eq:zdisp}) to the derived data points of $\sigma_{\Delta\redshift/\left(1+\redshift_{\mathrm{BCG}}\right)}({\rich}_{i})$ over all richness bins. 
We show the best fit of equation~(\ref{eq:zdisp}) together with the data points in the left panel of Figure~\ref{fig:photoz}, with the best-fit parameters, 
\[
(\delta_{\redshift}, \Gamma_{\redshift}) = (0.0093\pm0.0002, -0.18\pm0.05) \, .
\]
Note that we use $1165$, out of the $3057$ CAMIRA clusters, that have BCGs with available spectroscopic redshifts $\redshift_{\mathrm{BCG}}$ to determine equation~(\ref{eq:zdisp}).
Note that the usage of the functional form in equation~(\ref{eq:zdisp}) is supported by \citet[][see their Figure 5]{sereno15}, in which they studied a sample of clusters with a much larger size and found that the dispersion $\sigma_{\Delta\redshift}$ indeed distributes as a power-law function of richness.

On the other hand, we do not observe a monotonic redshift dependence in $\sigma_{\Delta\redshift/\left(1+\redshift_{\mathrm{BCG}}\right)}$ as seen in the right panel of Figure~\ref{fig:photoz}.
Rather, the value is roughly a constant, $\sigma_{\Delta\redshift/(1 + z_{\rm{BCG}})}\approx0.009$,
and shows a dip at $\redshift\approx0.4$.
This is in great agreement with \cite{murata19}, where larger dispersion at both low ($\redshift\lesssim0.4$) and high  ($\redshift\gtrsim0.6$) redshifts was seen than that at $\redshift\approx0.45$.
The larger dispersion at high redshift is expected, because photometry measurements of distant galaxies are noisier.
Meanwhile, larger dispersion at low redshift is mainly due to the lack of $u$-band data, as well as a difficulty in estimating the accurate color of close galaxies that are sometimes too bright for the HSC survey \citep{murata19}. 
Additionally, high-richness clusters are more abundant at low redshift than at high redshift (see Figure~\ref{fig:richz}), which could result in a redshift dependence in the best-fit parameters of $(\delta_{\redshift}, \Gamma_{\redshift})$.
Ideally, the photo-\redshift\ dispersion should be modeled as a function of both richness and redshift.
In this work, however, we cannot simultaneously constrain the richness- and redshift-dependence of the photo-\redshift\ dispersion, due to the lack of a large spec-\redshift\ sample.
We thus ignore the redshift dependence of $\sigma_{\Delta\redshift/\left(1+\redshift_{\mathrm{BCG}}\right)}$ in this work.
The number-weighted average of the richness- and redshift-dependent $\sigma_{\Delta\redshift/(1 + z_{\rm{BCG}})}$ over the whole sample is 
$0.0092$ and 
$0.0098$, respectively.
To first-order approximation, this corresponds to an increase at a level of 
0.0006 ($=0.0098 - 0.0092$)
if accounting for the redshift dependence in $\sigma_{\Delta\redshift/(1 + z_{\rm{BCG}})}$. 
A larger spec-\redshift\ sample is clearly warranted for future work with a detailed modeling of  $\sigma_{\Delta\redshift/\left(1+\redshift_{\mathrm{BCG}}\right)}$.

For the CMASS sample, the term $\sigma_{\Delta\redshift}$ is a subdominant factor, given that their redshifts are secured by spectroscopic observations with negligible measurement uncertainties.
We thus ignore the measurement uncertainty of the redshift of CMASS galaxies.

\subsection{Modeling of Halo Bias}
\label{sec:modelhalobias}

In this work, we model the halo bias of CMASS galaxies by a free parameter, $b_{\rm g}=b_{\mathrm{CMASS}}$.
This approach is identical to that in \cite{chuang13} and is sufficient for CMASS galaxies, in which the redshift-dependence of halo bias is not expected at this narrow range of redshift \citep{guo13}.
Meanwhile, the halo bias of CAMIRA clusters, $b_{\rm c}$, is linked to their cluster mass based on the \citet{tinker10} fitting formula.
Additionally, we account for the projection effect on the halo bias of CAMIRA clusters by following the method introduced in \cite{baxter16}.
In what follows, we briefly describe the modeling of the projection effect and refer interested readers to \cite{baxter16} for more details.

The projection effect is referred to that a single cluster is due to a projection of multiple systems aligning on the same sky position.
We only consider the projection of two halos because the projection of more than two systems is extremely rare and is therefore subdominant \citep{baxter16}.
Specifically, the halo bias \biash\ of a cluster at redshift \redshift\ with an observed richness \rich\ is modeled by an effective halo bias \bmodel, as a linear combination of the halo bias without projection, $\bnonproj(\rich,\redshift)$, and that due to projection, $\bproj(\rich,\redshift)$.
That is, 
\begin{equation}
\label{eq:bmodel}
\bmodel(\rich,\redshift) = (1-f) \bnonproj(\rich,\redshift) + f \bproj(\rich,\redshift) \, ,
\end{equation}
where $f$ is the probability that the cluster is a product of a projection effect\footnote{The parameter $f$ needs to be distinguished from the growth rate parameter $f_{\rm m}$.}.
The non-projected bias $\bnonproj(\rich,\redshift)$ is expressed as 
\begin{equation}
\label{eq:bnonproj}
\bnonproj(\rich,\redshift)= \frac{
\int \btinker(\Mfiveoo, \redshift)P( \rich|\Mfiveoo, \redshift, \mathbf{p}_{\mathrm{sr}}) N_{\mathrm{M}}( \Mfiveoo, \redshift) \dif\Mfiveoo
}{
\int P( \rich|\Mfiveoo, \redshift, \mathbf{p}_{\mathrm{sr}}) N_{\mathrm{M}}( \Mfiveoo, \redshift) \dif\Mfiveoo
} \, ,
\end{equation}
where $\btinker$ is the \citet{tinker10} halo bias, $N_{\mathrm{M}}(\Mfiveoo, \redshift)$ is the mass function evaluated by using the fitting formula in \cite{bocquet16}, and the term $P(\rich|\Mfiveoo, \redshift, \mathbf{p}_{\mathrm{sr}})$ with the \rtm\ parameters $\mathbf{p}_{\mathrm{sr}}$ describes the probability of observing the richness \rich\ given the cluster true mass \Mfiveoo\ at the redshift \redshift.
We stress that the form of equation~(\ref{eq:bnonproj}) already accounts for the Eddington and Malmquist bias, as proved and widely used in previous work \citep[e.g.,][]{liu15a, chiu16c, chiu18a,bulbul19,chiu20}.

It is important to note that the information of the \rtm\ scaling relation is fully contained in the term $P(\rich|\Mfiveoo, \redshift, \mathbf{p}_{\mathrm{sr}})$, which includes both measurement uncertainties and intrinsic scatter of richness at fixed cluster mass.
The \rtm\ scaling relation is characterized by the parameter $\mathbf{p}_{\mathrm{sr}} = \left( \Arich, \Brich, \Crich, \Drich \right)$ as
\begin{equation}
\label{eq:richness_to_mass}
\left\langle\ln\rich|\Mfiveoo\right\rangle = \ln\Arich + \Brich\ln\left( \frac{\Mfiveoo}{\MPIV} \right) + \Crich\ln\left(\frac{1 + \redshift}{1 + \ZPIV}\right) \, ,
\end{equation}
with log-normal intrinsic scatter \Drich\ at fixed mass,
where \Brich\ and \Crich\ are the mass and redshift power-law indices, respectively; 
\Arich\ is the normalization at the pivotal mass $\MPIV=10^{14}h^{-1}\Msun$ and the pivotal redshift $\ZPIV =0.6$.

As for the projected halo bias \bproj, it reads
\begin{equation}
\label{eq:bproj}
\bproj(\rich,\redshift) = \frac{
\int \btinker(\Mfiveootilde, \redshift)P( q \rich|\Mfiveoo, \redshift, \mathbf{p}_{\mathrm{sr}}) N( \Mfiveoo, \redshift) \dif\Mfiveoo
}{
\int P( q \rich|\Mfiveoo, \redshift, \mathbf{p}_{\mathrm{sr}}) N( \Mfiveoo, \redshift) \dif\Mfiveoo
} \, ,
\end{equation}
in which tilde put on mass $M$ stands for 
\begin{equation}
\label{eq:bmodified}
\tilde{M}=M\left[ 1 + g \left(\frac{1-q}{q}\right)^{\frac{1}{\Brich}}\right]
\end{equation}
where $g$ and $q$ are two nuisance parameters 
over the ranges
of $0\leq g \leq 1$ and $0.5 \leq q \leq 1$, respectively.

The scenario described by equation~(\ref{eq:bmodified}) is as follows.
In a projected system consisting of two halos with mass $M_1$ and $M_2$, respectively, the halo bias of the projected system is confined to be between those inferred by $M_1$ and $M_1 + M_2$ with a definition of $M_1 > M_2$.
Then, it is easy to write the halo bias of a projected system as 
\begin{equation}
\label{eq:bmodified_derivation}
\btinker(\Mfiveootilde, \redshift) = \btinker(M_1 + g M_2, \redshift) \, ,
\end{equation}
where $g$ is the nuisance parameter over the range of $0\leq g \leq 1$.
Assuming that the projected system is observed with a richness \rich, in which 
the 
two halos
with mass $M_1$ and $M_2$ have the richness of ${\rich}_1 = q\rich$ and ${\rich}_2 = (1-q)\rich = \frac{(1-q)}{q}{\rich}_1$ with $0.5 \leq q \leq 1$, respectively, 
we can obtain equation~(\ref{eq:bmodified}) by substituting $M_2 = \left(\frac{1-q}{q}\right)^{\frac{1}{\Brich}} M_1$ in equation~(\ref{eq:bmodified_derivation}) using the relation of $\rich\sim M^{\Brich}$ or $M\sim \rich^{\frac{1}{\Brich}}$.
In a limit of either no projection effect ($q=1$), or zero separation of two halos in a projected system ($q < 1$ and $g=1$), equation~(\ref{eq:bproj}) reduces to equation~(\ref{eq:bnonproj}).

Despite quite complex forms of equation~(\ref{eq:bproj}) and (\ref{eq:bmodified}), the physical interpretation is rather straightforward:
In a two-halo projected system with an observed richness \rich, the projection effect leads to an increasing halo bias of the main sub-halo with respect to that without projection.
As a result, this is equivalent to a shift by a factor of $g \left(\frac{1-q}{q}\right)^{\frac{1}{\Brich}}$ in \btinker\ given the richness ${\rich}_{1}=q\rich$ of the main sub-halo in equation~(\ref{eq:bproj}).

There are two assumptions implicitly made in modeling the projection effect.
First, we assume that observed richness at fixed cluster mass has no dependence on redshift.
This assumption is supported by various weak-lensing studies, suggesting that the redshift-trend power-law index \Crich\ of richness at fixed cluster mass is indeed consistent with zero \citep[e.g., ][]{murata19, chiu20}.
Second, in the model of a two-halo projected system, we assume that these two halos are located at the same redshift \redshift\ with a negligible redshift separation.
This is because a typical redshift separation of subhalos is at $\Delta\redshift\approx0.02$, which is much smaller than the width of our redshift binning and thus is negligible \citep{baxter16}.

The modeling of the projection effect in a greater depth requires an end-to-end validation based on large simulations \citep[see e.g.,][]{costanzi19,sunayama2020}, which is not available for our CAMIRA cluster sample.
However, the recent study of \cite{sunayama2020} suggests that a red-sequence based cluster finder could result in a cluster sample that preferentially selects systems locating at filaments aligning along the line of sight.
This selection bias changes the underlying mass distribution of optically selected cluster samples and ultimately bias their clustering measurements.
This selection bias is not included in our current modeling of the projection effect, which only accounts for the mis-match between observed richness and underlying true halo mass.
We refer readers to Section~\ref{sec:selection} for more discussions about the selection bias suggested by \cite{sunayama2020}.

\subsection{Modeling of Correlation Functions}
\label{sec:modelingofcorrelationfunctions}

Based on the formulation presented in
Sections~\ref{sec:modeling_fog} to \ref{sec:modelhalobias}, we further express the three power spectra in equation~(\ref{eq:generalform}) for modeling \xicamira, \xicamiracmass, and \xicmass. 
They are explicitly given by
\begin{align}
\label{eq:power_cluster}
P_{\mathrm{cc}}\left(\mathbf{k}\right)
=&
P_{\mathrm{m}}\left(k\right)
\left( 
b_{\rm c}
+ 
\growthrte(z_{\rm c})
\mu^2
\right)^2
\exp
\left(
-k^2
\mu^2
\sigma_{\redshift\mathrm{LoS}}^2
\right)
\, , 
\\
\label{eq:power_modified_cross}
P_{\mathrm{cg}}\left(\mathbf{k}\right)
=
&P_{\mathrm{m}}\left(k\right)
\left( 
b_{\rm c} + 
\growthrte(z_{\rm c})
\mu^2
\right)
\left( 
b_{\rm{g}}
+ 
\growthrte(z_{\rm g})
\mu^2
\right) \nonumber \\
&\times\exp
\left[
-k^2
\mu^2
\left(
\frac{
\sigma_{\redshift\mathrm{LoS}}^2 + \sigma_{v\mathrm{LoS}}^2
}{2}
\right)
\right]
\, , 
\\
\label{eq:power_modified_boss}
P_{\mathrm{gg}}\left(\mathbf{k}\right)
=
&P_{\mathrm{m}}\left(k\right)
\left( 
b_{\mathrm{g}}
+ 
\growthrte(z_{\rm g})
\mu^2
\right)^2 
\exp
\left(
-k^2
\mu^2
\sigma_{v\mathrm{LoS}}^2
\right)
\, ,
\end{align}
where the nonlinear velocity dispersion of CAMIRA clusters and the photo-\redshift\ uncertainty of CMASS galaxies are ignored (see sections \ref{sec:modeling_fog} and \ref{sec:model_photoz}, respectively).
The growth rate $\growthrte=\growthrte(z)$ is evaluated at the median redshift of a given sample, $z_{\rm c}$ and $z_{\rm g}$ for CAMIRA clusters and CMASS galaxies, respectively. 

The nonlinear velocity dispersion of CMASS galaxies, $\sigma_{v\mathrm{LoS}}$, is computed as
\begin{equation}
\label{eq:peculiar_velocity_of_cmass}
\sigma_{v\mathrm{LoS}} = \frac{  \sigma_{v,\redshift_{\rm g}} \times \left(1 + \redshift_{\mathrm{g}} \right)  }{H(\redshift_{\mathrm{g}})} 
\, , 
\end{equation}
where the line-of-sight velocity dispersion $\sigma_{v,\redshift_{\rm g}}$ of CMASS galaxies is fixed to $310~\mathrm{km}/\mathrm{sec}$, as suggested by our mocks (see Section~\ref{sec:modeling_fog}).
Note that we fix $\redshift_{\mathrm{g}}$ to $0.57$, as the median redshift of the CMASS sample.

Given a sample of CAMIRA clusters with a set of observed richness $N$, $\sigma_{\redshift\mathrm{LoS}}$ is computed as in equation~(\ref{eq:model_photoz_2}),
\begin{equation}
\label{eq:dispersion_cluster}
\sigma_{\redshift\mathrm{LoS}} = \frac{c}{H\left(\redshift_{\rm c}\right)} \overline{\sigma_{\Delta\redshift}} \, .
\end{equation}
where $\overline{\sigma_{\Delta\redshift}}$ is evaluated as the mean value of $\sigma_{\Delta\redshift}(\rich)$ (i.e., equation~(\ref{eq:zdisp})) among the clusters in the sample, given a set of parameters $\left(\delta_{\redshift}, \Gamma_{\redshift}\right)$. We use the cluster photo-\redshift\ in evaluating $\overline{\sigma_{\Delta\redshift}}$. 

The linear halo bias of CMASS galaxies, $b_{\rm g}=b_{\mathrm{CMASS}}$, is the only free parameter to model the amplitude of \xicmass. 
The remaining quantity is the halo bias of CAMIRA clusters, $b_{\rm c}$,
which is linked to the cluster mass and further connected to the observable (i.e., richness) by the \rtm\ scaling relation. In this way, one can calibrate the \rtm\ parameters by forward-modeling to an observed correlation function.
This process is referred to as the ``self-calibration'' of the \rtm\ relation based on clustering alone.
The halo bias of CAMIRA clusters is modeled as the mean value of the cluster sample, namely,
$b_{\rm c} = \overline{b_{\mathrm{model}}}$, where $\overline{b_{\mathrm{model}}}$ is the mean value of 
$b_{\mathrm{model}}\left(N, z\right)$ given by equation~(\ref{eq:bmodel}) over the cluster sample, given a set of parameters $\left(\Arich, \Brich, \Crich, \Drich, f, g, q\right)$.
It is worth mentioning that the we use the same cluster halo bias $b_{\rm c}$ in modelling both \xicamira\ and \xicamiracmass, while the clustering signals of the latter are dominated by the cluster-galaxy pairs at the range of overlapping redshift ($0.4\lesssim\redshift\lesssim0.7$).
We have verified that using a subsample of CAMIRA clusters, which are selected based on their redshifts such that the redshift distribution of clusters follows that of CMASS galaxies, to calculate $b_{\rm c}$ in modelling \xicamiracmass\ results in negligible difference.

To sum up, we will have nine free parameters, $\mathbf{p} = \left\lbrace\Arich, \Brich, \Crich, \Drich, f, g, q, \delta_{\redshift}, \Gamma_{\redshift} \right\rbrace$, in modeling \xicamira.
The first fourth parameters characterize the \rtm\ relation.
The parameters of $f$, $g$, and $q$ are used to model the projection effect, while the last two ($\delta_{\redshift}$ and $\Gamma_{\redshift}$) describe the redshift-smearing effect due to the photo-\redshift\ uncertainty of CAMIRA clusters.
The self-calibration of the \rtm\ relation is performed by modeling \xicamiracmass\ with an additional parameter of the CMASS halo bias $b_{\mathrm{CMASS}}$ on top of the nine free parameters above, and thus we have ten free parameters for the modeling of the cross-correlation function.

\subsection{Statistical Inference}
\label{sec:mcmc}

In this subsection, we describe the forward-modeling approach to calibrate the \rtm\ relation by modeling the measurements of auto- and/or cross-correlation functions in a framework of fixed cosmology.

We explore the parameter space using \texttt{emcee} \citep{foreman13,foreman19}, which implements the Affine Invariant Markov Chain Monte Carlo (MCMC) algorithm.
For a given data vector $\mathfrak{D}$ and the parameter vector $\mathbf{p}$, the posterior $\mathfrak{P}(\mathbf{p}|\mathfrak{D})$ of $\mathbf{p}$ is expressed as 
\begin{equation}
\label{eq:posteriors}
\mathfrak{P}(\mathbf{p}|\mathfrak{D}) = \mathcal{L(\mathfrak{D}|\mathbf{p})}\times\mathcal{P(\mathbf{p})} \, ,
\end{equation}
where $\mathcal{P}$ is the prior on $\mathbf{p}$, and $\mathcal{L(\mathfrak{D}|\mathbf{p})}$ is the likelihood of the model evaluated with the parameter vector $\mathbf{p}$.
The log-likelihood $\ln\mathcal{L(\mathfrak{D}|\mathbf{p})}$ reads
\begin{equation}
\label{eq:like}
\ln\mathcal{L(\mathfrak{D}|\mathbf{p})}= -\frac{1}{2} 
\left(
\mathfrak{M(\mathbf{p})} - \mathfrak{D} 
\right)^{\mathrm{T}}
\mathfrak{C}^{-1}
\left(
\mathfrak{M(\mathbf{p})} - \mathfrak{D} 
\right) 
\, ,
\end{equation}
where $\mathfrak{C}$ is the covariance matrix defined in Section~\ref{sec:covar}, and $\mathfrak{M}$ is the model of auto- and/or cross-correlation functions corresponding to the data vector $\mathfrak{D}$ (see Section~\ref{sec:modeling}).
The matter power spectra $P_{\mathrm{m}}(k)$ in equations~(\ref{eq:power_cluster}) and (\ref{eq:power_modified_cross}) are evaluated at the mean redshift of CAMIRA clusters in the sample (or the subsample).

Note that $\mathfrak{D}$ represents a generic term of data vectors, which can be a combination of various auto- and cross-correlation functions measured in the whole sample or different subsamples of richness and redshift.
In this work, we perform the modeling of \xicamira, \xicamiracmass, and \xicmass\ separately, and the joint modeling of $\xicamira+\xicamiracmass$ and $\xicamira + \xicamiracmass + \xicmass$.
For fitting the \xicamira\ alone, we have nine free parameters, $\mathbf{p} = \left\lbrace\Arich, \Brich, \Crich, \Drich, f, g, q, \delta_{\redshift}, \Gamma_{\redshift}\right\rbrace$.
For modeling \xicmass\ alone, as the simplest case, there is only one free parameter, the halo bias $b_{\rm{CMASS}}$.
If a cross-correlation function (\xicamiracmass) is included in the modeling, then we have ten free parameters, 
$\mathbf{p} = \left\lbrace\Arich, \Brich, \Crich, \Drich, f, g, q, \delta_{\redshift}, \Gamma_{\redshift}, b_{\mathrm{CMASS}}\right\rbrace$.

In this work, we cannot meaningfully constrain all parameters at the same time without informative priors, given current measurement uncertainties as well as the degeneracy among parameters.
Therefore, we focus on constraining the normalization \Arich\ of the \rtm\ relation while applying the informative priors on other parameters.
Specifically, we apply Gaussian priors of $\mathcal{N}(0.92, 0.13^2)$, $\mathcal{N}(-0.48, 0.69^2)$, and $\mathcal{N}(0.15, 0.07^2)$ on \Brich, \Crich, and \Drich, respectively.
These priors are suggested by the posteriors independently constrained by lensing magnification \citep{chiu20} and are statistically consistent with those obtained by a joint analysis of weak shear and cluster abundance \citep{murata19}.
Only a uniform prior between $0$ and $100$ is applied on \Arich.

A Gaussian prior $\mathcal{N}(0.1, 0.05^2)$ is applied on the parameter $f$ with an additional requirement of $0\leq f\leq 1$ to describe the percentage of projected systems in the sample.
This value is suggested by another optically selected cluster sample in the SDSS \citep{baxter16,simet17}.
Flat priors of $\mathcal{U}(0,1)$ and $\mathcal{U}(0.5,1)$ are applied on $g$ and $q$, respectively.
Note that we cannot well constrain the parameters of $f$, $g$, and $q$ based on cluster clustering alone, for which a dedicated effort using large simulations is needed \citep[e.g.,][]{costanzi19} and is currently not available for our sample.
By applying these priors in a MCMC framework, we effectively marginalize these parameters over the range of the parameter space with a minimal requirement of informative knowledge.
For the parameters of $\delta_{\redshift}$ and $\Gamma_{\redshift}$, 
we apply Gaussian priors of $\mathcal{N}(0.0093, 0.0002^2)$ and $\mathcal{N}(-0.18, 0.05^2)$, which are suggested by our data (see Section~\ref{sec:model_photoz}), respectively.

The halo bias $b_{\mathrm{CMASS}}$ is varied with a Gaussian prior $\mathcal{N}(1.93, 0.17^2)$, which was the posterior independently constrained by RSD and BAO together with the \WMAP9 CMB data \citep{chuang13}.
Although different cosmological parameters are used and fixed in this work, we note that the constraint of $b_{\mathrm{CMASS}}$ in \cite{chuang13} takes into account the variation of cosmological parameters and serves an adequate prior here.
We stress that the strategy in this work is to leverage the clustering of CMASS galaxies to improve the constraints on the mass calibration of CAMIRA clusters; therefore, imposing an informative prior on $b_{\mathrm{CMASS}}$ from an BAO analysis in the modelling of \xicamiracmass\ is a reasonable approach.
However, we note that the Gaussian prior on $b_{\mathrm{CMASS}}$ is not informative, once the measurement of \xicmass\ is included in the modelling.
This is because the constraining power on $b_{\mathrm{CMASS}}$ based on \xicmass\ alone is significantly stronger than the imposed prior.
In this case, removing the informative prior on $b_{\mathrm{CMASS}}$ results in negligible difference to that with the prior.
In the interest of a uniform analysis in this work, however, we still consistently apply the Gaussian prior on $b_{\mathrm{CMASS}}$ in all modelling, even those including \xicmass.
A summary of the adopted priors is given in Table~\ref{tab:priors}.

\begin{table}
\centering
\begin{tabular}{ccc}
\hline\hline
Parameter &Priors &Reference\\[3pt]
\hline
\multicolumn{3}{c}{Mass calibration} \\
\hline
\Arich &$\mathcal{U}(0,100)$ &Section~\ref{sec:mcmc} \\[3pt]
\Brich &$\mathcal{N}(0.92, 0.13^2)$  &\cite{chiu20} \\[3pt]
\Crich &$\mathcal{N}(-0.48, 0.69^2)$  &\cite{chiu20} \\[3pt]
\Drich &$\mathcal{N}(0.15, 0.07^2)$    &\cite{chiu20} \\[3pt]
\hline
\multicolumn{3}{c}{Projection effect} \\
\hline
$f$ &$\mathcal{N}(0.1, 0.05^2)$ and $0\leq f\leq1$ &\cite{baxter16} \\[3pt]
$g$ &$\mathcal{U}(0, 1^2)$ &\cite{baxter16} \\[3pt]
$q$ &$\mathcal{U}(0.5, 1^2)$ &\cite{baxter16} \\[3pt]
\hline
\multicolumn{3}{c}{Photo-\redshift\ uncertainty} \\
\hline
$\delta_{\redshift}$ &$\mathcal{N}(0.0093, 0.0002^2)$ &Section~\ref{sec:model_photoz} \\[3pt]
$\Gamma_{\redshift}$ &$\mathcal{N}(-0.18, 0.05^2)$ &Section~\ref{sec:model_photoz} \\[3pt]
\hline
\multicolumn{3}{c}{Halo bias of CMASS galaxies} \\
\hline
$b_{\mathrm{CMASS}}$ &$\mathcal{N}(1.93, 0.17^2)$ &\cite{chuang13} \\[3pt]
\hline\hline
\end{tabular}
\caption{
Summary of the adopted priors used in the modeling.
The first column represents the names of the parameters.
The second column describes the priors used in the modeling, while the last column states the references to the informative priors.
}
\label{tab:priors}
\end{table}
\begin{figure*}
\centering
\resizebox{\textwidth}{!}{
\includegraphics[scale=1]{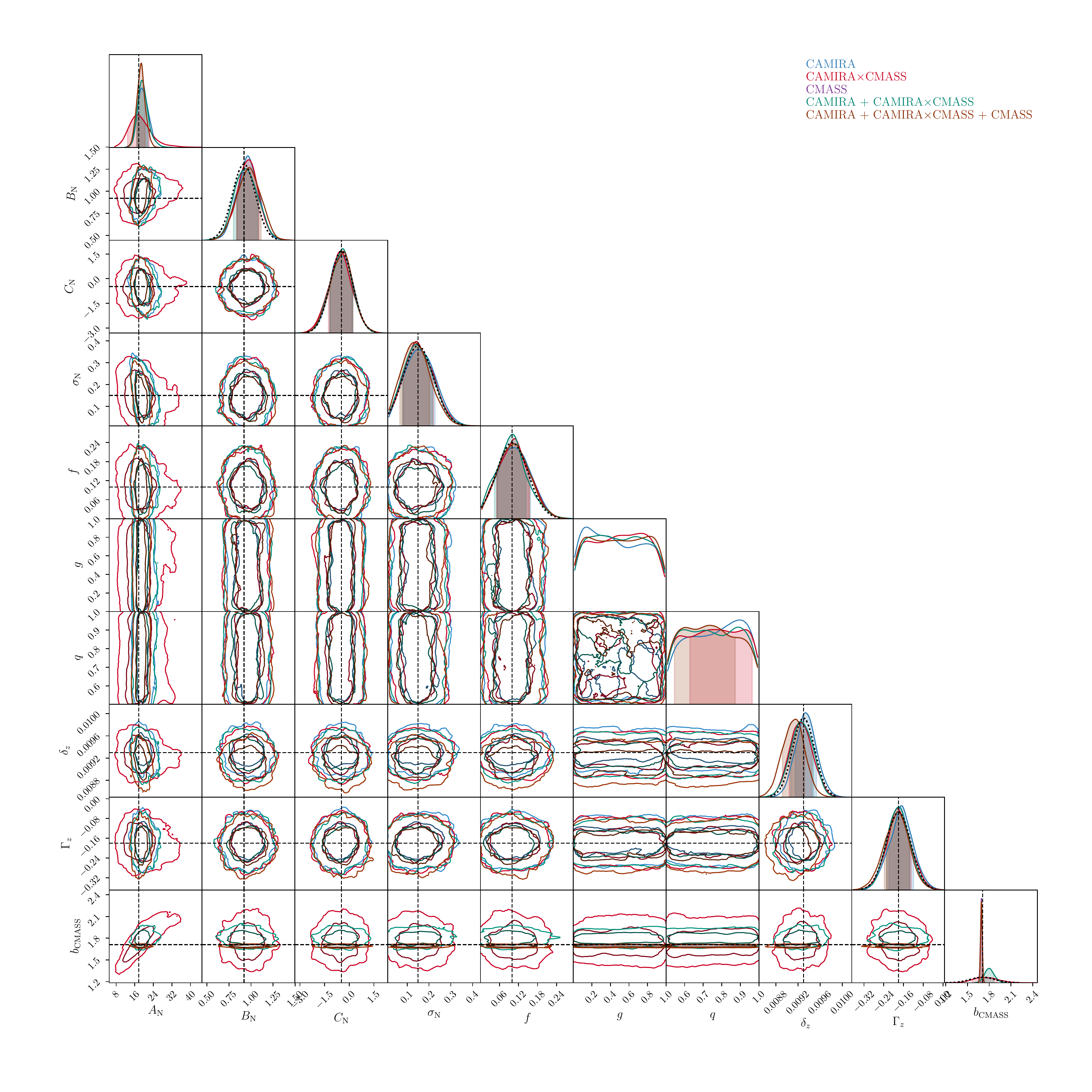}
}
\vspace{-1.5cm}
\caption{
Constraints on the parameters in the mock validations.
The constraints obtained based on different data sets are marked by different colors, as shown on the upper-right corner.
These contours are smoothed through a kernel density estimation, while the shaded regions represent the $68\%$ confidence intervals of the parameters.
The input values of the \rtm\ relation parameters are indicated by the dashed lines associating with the constraints of $(\Arich, \Brich, \Crich, \Drich)$.
The dashed lines associated with the parameters $(\delta_{\redshift}, \Gamma_{\redshift})$ indicate the constraints on the richness-dependent dispersion of the photo-\redshift\ uncertainty of CAMIRA clusters obtained from the data, as the input values for creating mock catalogs.
The dashed lines associated with the parameters of $f$, $g$, and $q$, which characterize the projection effect on the halo bias of clusters, indicate the means of the adopted Gaussian priors.
The dashed line associated with the linear halo bias of mock CMASS galaxies indicate the median value of $b_{\mathrm{CMASS}}$ evaluated based on the true halo mass following the \citet{tinker10} formula.
The dotted lines in the posterior panels of parameters indicate the adopted Gaussian priors (see Table~\ref{tab:priors}).
}
\label{fig:triangle_plots_3dmocktests}
\end{figure*}
\begin{figure}
\centering
\resizebox{0.5\textwidth}{!}{
\includegraphics[scale=1]{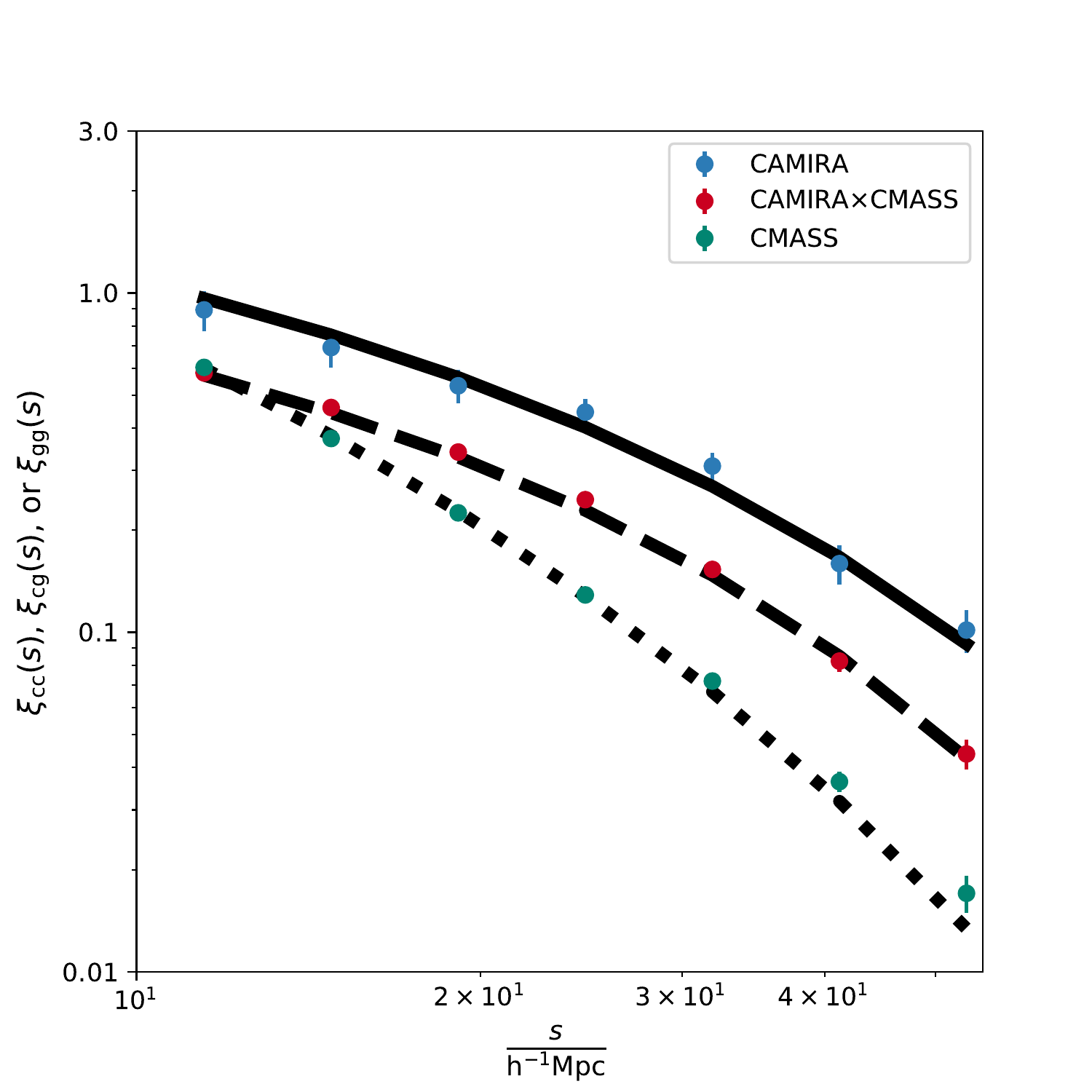}
}
\caption{
The best-fit correlation functions and the measurements in the mock validations.
The stacked measurements of \xicamira, \xicamiracmass, and \xicmass\ of 10 mock data sets are shown as the blue, red, and green circles, respectively.
Meanwhile, the best-fit correlation functions of \xicamira, \xicamiracmass, and \xicmass\ are shown by the solid, dashed, and dotted lines, respectively.
The solid line is generated using the best-fit parameters obtained from modeling \xicamira\ alone (blue in Figure~\ref{fig:triangle_plots_3dmocktests}), while the dashed and dotted lines are obtained from the joint modeling of $\xicamira + \xicamiracmass + \xicmass$ (brown in Figure~\ref{fig:triangle_plots_3dmocktests}).
}
\label{fig:mockprofiles}
\end{figure}

\subsection{Validations using mock catalogs}
\label{sec:validation}

By using the mock catalogs, we perform end-to-end validation tests of our codes and the assumptions made in the modeling.
For example, in this work we assume that the power spectrum can be simply described by equation~(\ref{eq:generalform}), which only models the effects of the RSD, FoG, and photo-\redshift\ smearing without accounting for, e.g., the assembly bias \citep{lin16, zu17}.
By carrying out the modeling on mock measurements in the identical way as on the data, our goal is to ensure that we can recover the input parameters $(\Arich, \Brich, \Crich, \Drich)$ of the \rtm\ scaling relation without significant bias.

To do so, we randomly draw 10 different sets of mock measurements to enlarge the sample size in mock modeling, such that the bias (if exists) would not be hidden by statistical uncertainties, which are $3.2\left(\approx\sqrt{10}\right)$ times smaller than those of the observations.
For a combination of data vectors in equation~(\ref{eq:posteriors}), the posteriors of the parameters in the joint modeling of the mock measurements are
\begin{equation}
\label{eq:posteriors_mocks}
\mathfrak{P}(\mathbf{p}|\left\lbrace
\mathfrak{D}_{1}, \mathfrak{D}_{2}, \cdots, \mathfrak{D}_{10}  
\right\rbrace) = 
\left[
\prod_{i=1}^{10}\mathcal{L(\mathfrak{D_{i}}|\mathbf{p})}
\right]
\times\mathcal{P(\mathbf{p})} \, ,
\end{equation}
where $\mathfrak{D_{i}}$ is the $i$-th mock measurement.

Despite the efforts in carefully mimicking observational properties of CAMIRA  clusters and CMASS galaxies in the mock catalogs, 
we note some of the limitations of the mocks.
First, the redshift estimates of mock catalogs only contain measurement uncertainties, distributed as a Gaussian distribution, but not the systematics introduced by the CAMIRA cluster finder.
This inevitably results in a moderate discrepancy in the redshift distribution between the mocks and the real data.
Second, we change the mean of the Gaussian prior on $b_{\mathrm{CMASS}}$ from $1.93$ to $1.71$, which is the median value of the CMASS halo bias evaluated using the true halo mass following the \citet{tinker10} formula.
Given the systematic uncertainty in the halo bias of mock halos at a level of $10\percent$ (see Section~\ref{sec:mocks}), this value is statistically consistent ($\lesssim1\sigma$) with the observational constraint from \cite{chuang13}.
Third, we still simultaneously model the parameters of $(f, g, q)$ in the identical way as described in Section~\ref{sec:modelingofcorrelationfunctions}, although there is no projection effect on the halo bias of mock clusters. 
Fourth, there is no selection bias 
in mock cluster catalogs, as opposed to the real data.
It has been suggested that a selection bias could exist in a cluster sample constructed by a red-sequence based algorithm \citep{sunayama2020}, such as \texttt{CAMIRA}.
We refer readers to Section~\ref{sec:selection} for more discussions about this selection bias.
Fifth, the \rtm\ relation in mock cluster catalogs is assumed to have log-normal intrinsic scatter of richness at fixed mass.
It is worth mentioning that a redshift-dependent form for intrinsic scatter is suggested for CAMIRA clusters, although without statistically significant evidence \citep{murata19}.

The results of the mock validations are shown in Figure~\ref{fig:triangle_plots_3dmocktests}, where we show the constraints of the parameters from 
the modeling of \xicamira, \xicamiracmass, and \xicmass, independently, and the combination among them.
As seen in Figure~\ref{fig:triangle_plots_3dmocktests}, our modeling can recover the input parameters (within $1\sigma$), indicated by the dashed lines.
This suggests that (1) our modeling approach can deliver an unbiased result, and that (2) the assumptions made in constructing the models are valid.
We further show the best-fit profiles and the mock measurements in Figure~\ref{fig:mockprofiles}, demonstrating that our models provide a good description of the measurements.
Note that the mock validations on the subsample analysis deliver the same picture as the default analysis, we thus only present the results of mock validations using the cluster sample as a whole.

It is interesting to note that the constraint on the normalization \Arich\ in modeling \xicamiracmass\ alone is weaker than that in modeling \xicamira\ alone.
This is due to a strong degeneracy between the normalization \Arich\ and the CMASS halo bias $b_{\mathrm{CMASS}}$, which is completely dominated by the prior.
Consequently, the constraint on \Arich\ largely depends on the prior on $b_{\mathrm{CMASS}}$  and becomes weaker than that from modeling \xicamira\ alone.
On the other hand, the joint modeling of \xicamira\ and \xicamiracmass\ (shown in green) shows that the constraint on \Arich\ completely follows that in the modeling of \xicamira\ alone (shown in blue), effectively breaking the \Arich-$b_{\mathrm{CMASS}}$ degeneracy.
Meanwhile, the inclusion of \xicmass\ in the joint modeling (shown in brown) essentially pins down the CMASS halo bias $b_{\mathrm{CMASS}}$, resulting in an absolute calibration of \Arich\ with slightly better accuracy and precision.

To sum up, we conclude that our modeling strategies can recover the underlying true values of the parameters within uncertainties, as suggested by our end-to-end mock validations.

\begin{figure*}
\centering
\resizebox{\textwidth}{!}{
\includegraphics[scale=1]{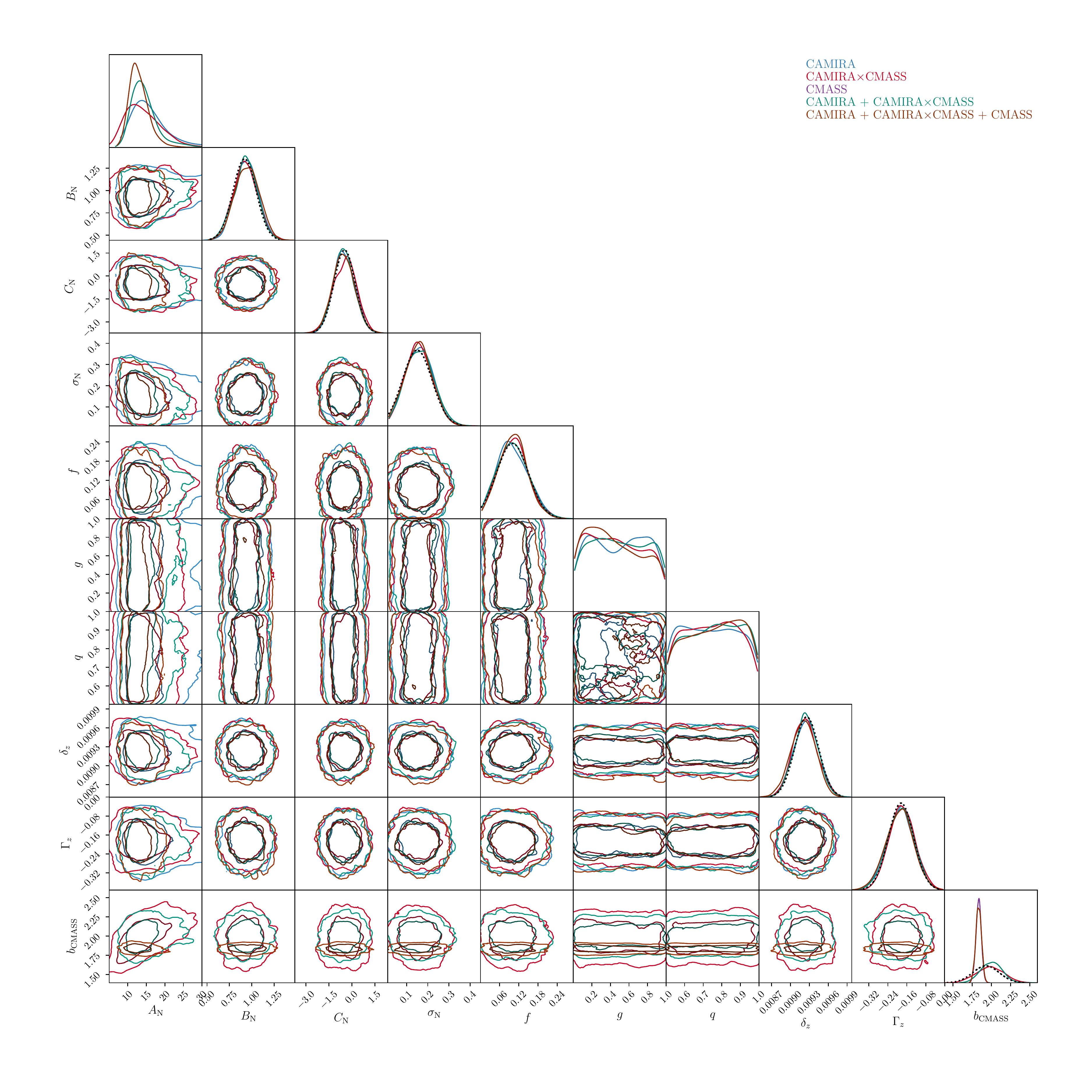}
}\vspace{-1.2cm}
\caption{
Parameter constraints obtained in the modeling of observed correlation functions.
This plot is generated in the same configuration as in Figure~\ref{fig:triangle_plots_3dmocktests}, except that the results are obtained by modeling the observed \xicamira, \xicamiracmass, \xicmass, or combinations of them.
Note that, for clarity, we only show the constraints based on the default analysis with CAMIRA clusters as a whole, as the subsample analysis delivers a consistent result (see Figure~\ref{fig:triangle_plots_subsamples}).
}
\label{fig:triangle_plots}
\end{figure*}
\begin{table*}
    \centering
    \resizebox{\textwidth}{!}{    
    \begin{tabular}{lccccccc}
        \hline\hline
		Data sets & $A_{\rm{N}}$ & $B_{\rm{N}}$ & $C_{\rm{N}}$ & $\sigma_{\rm{N}}$ & $\delta_{z}$ & $\Gamma_{z}$ & $b_{\rm{CMASS}}$ \\[4pt]
        \hline
		\multicolumn{8}{c}{Default analysis (no binning in clusters)} \\[4pt]
		\hline
CAMIRA & $13.8^{+5.8}_{-4.2}$ & $0.95^{+0.12}_{-0.14}$ & $-0.67^{+0.84}_{-0.55}$ & $0.165^{+0.061}_{-0.072}$ & $\left( 92.6^{+1.9}_{-2.2} \right) \times 10^{-4}$ & $-0.183^{+0.053}_{-0.063}$ & -- \\[2pt]
CAMIRA$\times$CMASS & $11.8^{+6.0}_{-3.5}$ & $0.91^{+0.15}_{-0.12}$ & $-0.30^{+0.62}_{-0.92}$ & $0.149^{+0.069}_{-0.061}$ & $\left( 92.4^{+1.7}_{-2.2} \right) \times 10^{-4}$ & $-0.186^{+0.061}_{-0.050}$ & $1.98^{+0.19}_{-0.18}$ \\[2pt]
CMASS & -- & -- & -- & -- & -- & -- & $1.838^{+0.032}_{-0.038}$ \\[2pt]
CAMIRA + CAMIRA$\times$CMASS & $13.2^{+3.4}_{-2.7}$ & $0.93^{+0.15}_{-0.12}$ & $-0.59^{+0.71}_{-0.67}$ & $0.176^{+0.050}_{-0.087}$ & $\left( 92.3^{+2.2}_{-1.6} \right) \times 10^{-4}$ & $-0.178^{+0.050}_{-0.064}$ & $2.02^{+0.12}_{-0.16}$ \\[2pt]
CAMIRA + CAMIRA$\times$CMASS + CMASS & $11.9^{+3.0}_{-1.9}$ & $0.98^{+0.11}_{-0.17}$ & $-0.62^{+0.78}_{-0.66}$ & $0.164^{+0.056}_{-0.077}$ & $\left( 92.2^{+1.9}_{-2.1} \right) \times 10^{-4}$ & $-0.178^{+0.046}_{-0.070}$ & $1.826^{+0.045}_{-0.029}$ \\[2pt]
        \hline
        \multicolumn{8}{c}{Subsample analysis} \\[4pt]
        \hline
CAMIRA  & $14.3^{+5.6}_{-3.8}$ & $0.94^{+0.15}_{-0.12}$ & $-0.41^{+0.63}_{-0.65}$ & $0.158^{+0.061}_{-0.071}$ & $\left( 92.6\pm 1.9 \right) \times 10^{-4}$ & $-0.190^{+0.061}_{-0.055}$ & -- \\[2pt]
CAMIRA$\times$CMASS & $12.2^{+4.0}_{-3.6}$ & $0.89^{+0.15}_{-0.11}$ & $-0.62^{+0.80}_{-0.60}$ & $0.158^{+0.064}_{-0.080}$ & $\left( 92.0^{+2.2}_{-1.8} \right) \times 10^{-4}$ & $-0.189^{+0.054}_{-0.053}$ & $2.00^{+0.13}_{-0.22}$ \\[2pt]
CAMIRA + CAMIRA$\times$CMASS & $14.5^{+3.4}_{-2.7}$ & $0.91^{+0.15}_{-0.11}$ & $-0.32^{+0.66}_{-0.59}$ & $0.165^{+0.056}_{-0.081}$ & $\left( 92.7^{+1.5}_{-2.5} \right) \times 10^{-4}$ & $-0.188^{+0.053}_{-0.046}$ & $2.03^{+0.12}_{-0.13}$ \\[2pt]
CAMIRA + CAMIRA$\times$CMASS + CMASS & $12.2^{+2.6}_{-1.9}$ & $0.93^{+0.10}_{-0.14}$ & $-0.10^{+0.43}_{-0.73}$ & $0.167^{+0.056}_{-0.077}$ & $\left( 91.3^{+2.2}_{-1.6} \right) \times 10^{-4}$ & $-0.174^{+0.042}_{-0.070}$ & $1.849^{+0.032}_{-0.040}$ \\[2pt]
        \hline
        \multicolumn{8}{c}{Subsample analysis without the Gaussian priors on $B_{N}$, $C_{N}$} \\[4pt]
        \hline
CAMIRA & $16.0^{+9.8}_{-7.2}$ & $1.84^{+0.13}_{-0.59}$ & $0.2^{+3.1}_{-1.8}$ & $0.118^{+0.096}_{-0.054}$ & $\left( 92.6^{+2.0}_{-1.8} \right) \times 10^{-4}$ & $-0.180^{+0.052}_{-0.057}$ & -- \\[2pt]
CAMIRA$\times$CMASS & $11.5^{+6.8}_{-5.4}$ & $1.01^{+0.66}_{-0.39}$ & $-0.5^{+2.6}_{-2.5}$ & $0.142^{+0.079}_{-0.061}$ & $\left( 93.2^{+1.4}_{-2.6} \right) \times 10^{-4}$ & $-0.188^{+0.039}_{-0.066}$ & $1.96^{+0.20}_{-0.16}$ \\[2pt]
CAMIRA + CAMIRA$\times$CMASS & $14.7^{+7.0}_{-5.0}$ & $1.35^{+0.50}_{-0.31}$ & $-0.3^{+2.6}_{-1.5}$ & $0.154^{+0.070}_{-0.065}$ & $\left( 92.1^{+1.8}_{-2.0} \right) \times 10^{-4}$ & $-0.186^{+0.050}_{-0.051}$ & $2.07^{+0.11}_{-0.16}$ \\[2pt]
CAMIRA + CAMIRA$\times$CMASS + CMASS & $13.1^{+3.2}_{-3.6}$ & $1.24^{+0.65}_{-0.25}$ & $0.6^{+2.5}_{-1.5}$ & $0.147^{+0.068}_{-0.066}$ & $\left( 91.1^{+2.5}_{-1.6} \right) \times 10^{-4}$ & $-0.193^{+0.049}_{-0.059}$ & $1.845^{+0.032}_{-0.043}$ \\[2pt]
        \hline
    \end{tabular}
    }
\caption{
Parameter constraints obtained from the modeling of observed correlation functions.
The results of the default analysis with all clusters (i.e., without the binning in the cluster richness and redshift) are presented in the first tier,
while the constraints of the subsample analysis are shown in the second tier.
The third tier represents the constraints in the subsample analysis without the Gaussian priors on \Brich\ and \Crich.
The first column indicates the measurements used in the modeling.
The second to fifth columns present the constraints of the \rtm\ relation parameters, as defined in equation~(\ref{eq:richness_to_mass}).
The sixth and seventh columns are the parameters characterizing the richness-dependent dispersion of the redshift uncertainty of CAMIRA clusters (see equation~(\ref{eq:zdisp})).
The last column is the constraint on the linear halo bias of CMASS galaxies.
Note that, for clarity, the parameters characterizing the projection effect (i.e., $f$, $g$, $q$) are not shown, because they are strictly following the priors \citep[as also seen in][]{baxter16}.
}
\label{tab:params}
\end{table*}
%

%
%

\section{Results and discussion}
\label{sec:results}

In this section, we present and discuss the results of the self-calibration of the \rtm\ scaling relation by modeling the redshift-space auto/cross-correlation functions.
Given the data sets in this work, we can only constrain the normalization \Arich\ of the \rtm\ relation with informative priors applied on other parameters.
In addition, the modeling is validated against the tests of large mock catalogs, ensuring that our results are unbiased.

We first show the constraints of the parameters in Figure~\ref{fig:triangle_plots}, where the results based on different measurements are marked by different colors.
Note that, for clarity, Figure~\ref{fig:triangle_plots} only contains the constraints based on the default analysis with all clusters, as the subsample analysis essentially returns consistent results (within $1\sigma$).
The constraints obtained from the subsample analysis are presented in Figure~\ref{fig:triangle_plots_subsamples}.
These constraints are also tabulated in Table~\ref{tab:params}.
In Figure~\ref{fig:triangle_plots}, it is clear that the posteriors of the parameters, except \Arich\ and $b_{\mathrm{CMASS}}$, are all largely following the adopted priors, as expected from the mock validations (see Section~\ref{sec:validation}).
It is also worth mentioning that the correlation patterns among the parameters in Figure~\ref{fig:triangle_plots} are in great agreement with those based on the mock validations (see Figure~\ref{fig:triangle_plots_3dmocktests}), suggesting that the mock catalogs indeed well describe the observed properties of the CAMIRA and CMASS samples.

\begin{figure*}
\centering
\resizebox{0.48\textwidth}{!}{
\includegraphics[scale=1]{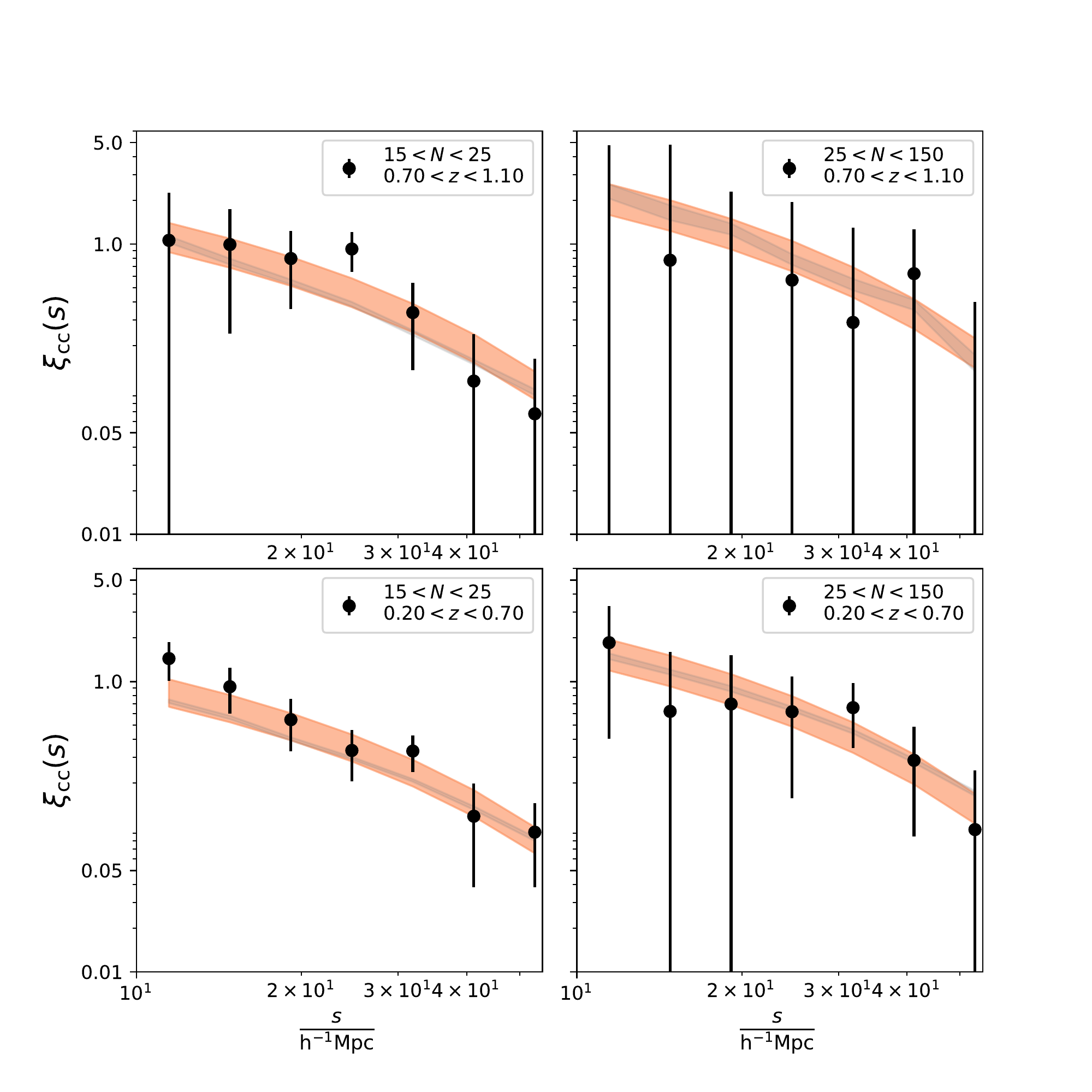}
}
\resizebox{0.48\textwidth}{!}{
\includegraphics[scale=1]{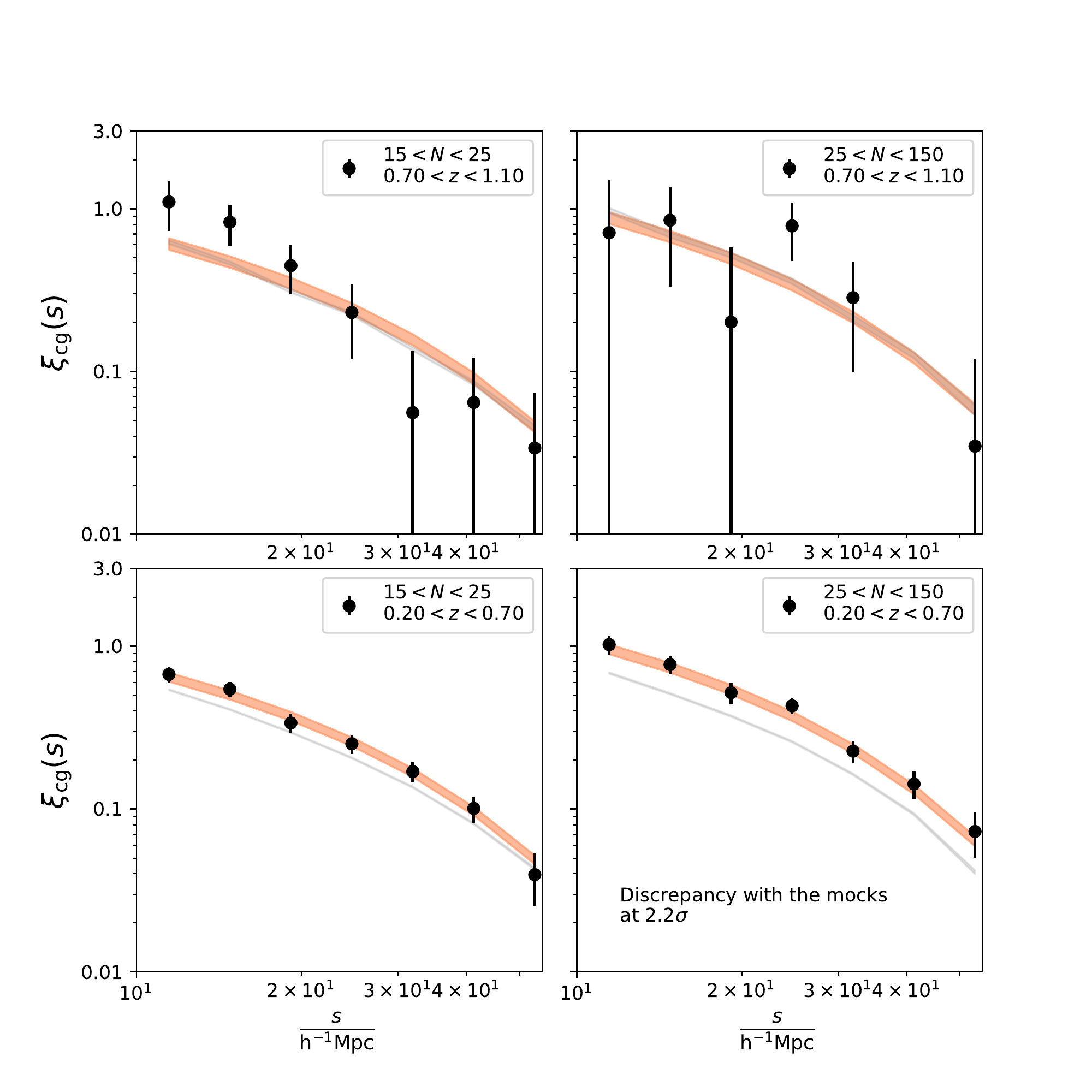}
}
\caption{
Left panels: Observed auto-correlation function \xicamira\ of CAMIRA clusters in different richness and redshift bins.
The best-fit models (evaluated using the constraints shown in the sixth rows in Table~\ref{tab:params}) with the $68\percent$ confidence regions are shown as the red regions.
The shaded grey regions are the mean of the auto-correlations among the 432 mock cluster catalogs.
Right panels: Observed cross-correlation \xicamiracmass\ between CAMIRA clusters and CMASS galaxies, the best-fit models, and the mean of those estimated from mock catalogs in the same manner as in the left panel.
The best-fit models of  \xicamiracmass\ are evaluated using the parameter constraints obtained from the joint modeling of CAMIRA+CAMIRA$\times$CMASS+CMASS (the ninth row in Table~\ref{tab:params}).
Additionally, we mark the discrepancy between mocks and observations in the lower corner of each subplot if it exceeds $2\sigma$.
}
\label{fig:auto_profile}
\end{figure*}

We then show the measurements (black points) and the best-fit profiles\footnote{These profiles are evaluated using the best-fit parameters in the sixth row in Table~\ref{tab:params}} (red regions) of the CAMIRA auto-correlation functions in the subsample analysis in the left panel of Figure~\ref{fig:auto_profile}.
In addition, the $68\%$ confidence regions of the mean of the correlation functions among the 432 mock catalogs are indicated by the grey shaded area.
Although the errorbars are large, it is seen that (1) the best-fit models provide a good description for the observed correlation functions, and that (2) the observed correlation functions show a hint for slightly higher amplitudes than those measured from the mocks (grey regions).

The discrepancy can be seen more clearly in the right panel of Figure~\ref{fig:auto_profile}, where we present the more precise measurements of the cross-correlation functions between the CAMIRA and the CMASS samples.
In this case, the best-fit models (red regions) are produced based on the joint modeling of \xicamira, \xicamiracmass, and \xicmass\ in the subsample analysis\footnote{The ninth row in Table~\ref{tab:params}}.
There indeed exists a discrepancy between the measured \xicamiracmass\ (black circles) and those estimated from the mocks (grey regions), especially the low-redshift sample at $0.2<\redshift<0.7$.
In addition, this discrepancy is higher for high-richness clusters (at the level of $2.2\sigma$) than the low-richness samples (at the level of  $1\sigma$)\footnote{Note that we take into account the correlation among the radial bins in calculating the significance of these discrepancies.}.
We expect that this discrepancy could be mitigated by accounting for the redshift dependence in the photo-\redshift\ dispersion of CAMIRA clusters (see Section~\ref{sec:modeling}).
This is because the redshift distribution of the CMASS sample is peaked at the redshift of $\redshift\approx0.5$, such that the resulting \xicamiracmass\ is weighted at this redshift, which is approximately the minimum of the photo-\redshift\ dispersion of CAMIRA clusters.
A smaller dispersion in the redshift uncertainty results in a higher amplitude of a correlation function, which is consistent with the observed \xicamiracmass\ as opposed to the mocks.
Taking into account the photo-\redshift\ dispersion as a decreasing function of richness, this discrepancy is expected to be larger in the high-richness bin than the low-richness bin, as also seen in our results.
Currently, this discrepancy is not significant in this work (i.e., at a level of $2.2\sigma$ for the high-richness and low-redshift bin, and $\approx1.2\sigma$ in the whole-sample analysis). 
This implies that modeling the photo-\redshift\ dispersion as a function of richness and redshift is required in improving the mock catalogs in the future.

With the precision in \xicamiracmass\ in the subsample analysis, we can constrain the mass- and redshift-trend power-law indices of the \rtm\ relation without the Gaussian priors.
Specifically, we replace the Gaussian priors $\mathcal{N}(0.92, 0.13^2)$ and $\mathcal{N}(-0.48, 0.69^2)$ on \Brich\ and \Crich\ with the uniform priors $\mathcal{U}(0, 2)$ and $\mathcal{U}(-5,5)$, respectively, and then repeat the whole modeling.
The resulting constraints are shown in Figure~\ref{fig:triangle_plots_without_prior} and are tabulated in Table~\ref{tab:params}.
Except for
the poor constraints obtained from the modeling based on \xicamira\ alone, it can be seen that the resulting constraints on \Arich, \Brich, and \Crich\ are all statistically consistent with those from the default analysis.
This suggests that the constraint on \Arich\ from the default analysis is not sensitive to the adopted Gaussian priors on \Brich\ and \Crich.

\begin{figure}
\centering
\resizebox{0.5\textwidth}{!}{
\includegraphics[scale=1]{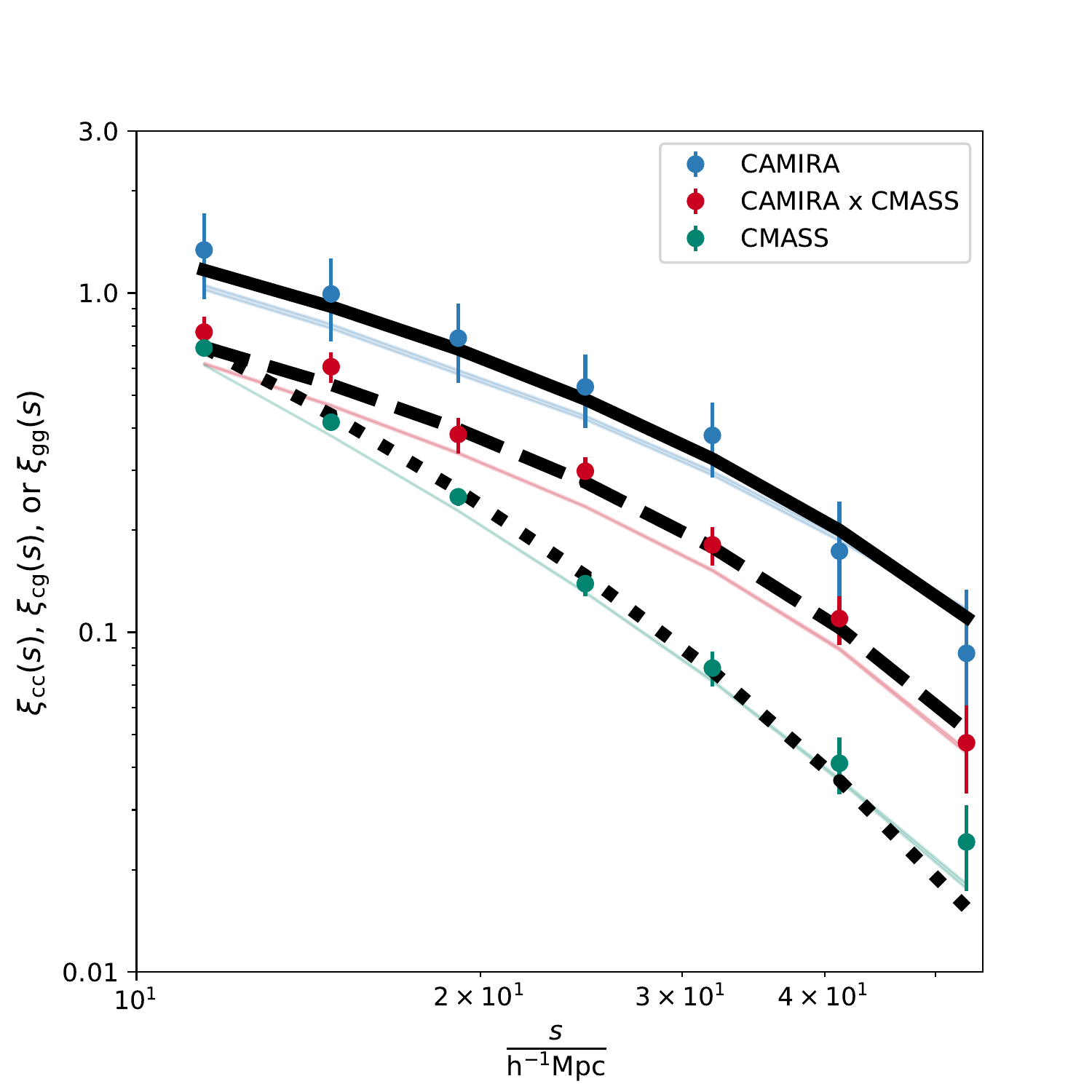}
}
\caption{
Auto- and cross-correlation functions in the default analysis.
The observed correlation functions (best-fit models) of \xicamira, \xicamiracmass, and \xicmass\ are shown by the blue, red, and green circles  (solid, dashed, and dotted lines), respectively.
The mean values of  \xicamira, \xicamiracmass, and \xicmass\  among the 432 mocks are marked by the blue, red, and green shade regions, respectively.
}
\label{fig:boss_profile}
\end{figure}

In Figure~\ref{fig:boss_profile}, we show the results of the default analysis (i.e., without binning the clusters in richness and redshift).
The observed \xicamira\ (blue circles) also shows a hint for a higher amplitude than the mean value of the 432 mocks (blue shaded region), although they are consistent with each other given the errorbars.
On the other hand, the observed \xicamiracmass\ (red circles) clearly shows a higher amplitude compared to the mocks (red shaded regions). 
This enhancement is at a level of $1.2\sigma$, accounting for the correlation among the radial bins.
Last, the observed \xicmass\ (green circles) is also higher than the mean value of the 432 mocks (green shaded regions) at a level of $2\sigma$.

To sum up, the observed correlation functions show higher amplitudes than those estimated from the mocks.
These discrepancies are at levels of $\lesssim0.5\sigma$, $1.2\sigma$ and $2\sigma$ for \xicamira, \xicamiracmass, and \xicmass, respectively.
In terms of \xicamiracmass, the discrepancy is mainly attributed to the low-redshift sample, especially for the high-richness clusters.
In addition, Figures~\ref{fig:auto_profile} and \ref{fig:boss_profile} show that these discrepancies are nearly independent of the scale, suggesting that the difference is due to the linear halo bias which changes the overall normalization.

While our constraint on the linear halo bias of the CMASS galaxies  ($b_{\mathrm{CMASS}} = 1.838^{+0.032}_{-0.038}$) obtained from the modeling of \xicmass\ alone is in good agreement with the independent result of \cite{chuang13}, $b_{\mathrm{CMASS}}=1.93\pm0.17$,  the observed amplitude of \xicmass\  is higher than that from the mocks at a level of $\approx10\percent$ (or $\approx2\sigma$).
This corresponds to a higher linear halo bias for the observed CMASS galaxies at a level of $\approx5\percent$.
Note that this comparison only accounts for the statistical uncertainty but not the systematics between the simulations and the observation.
Recalling that the distribution of the linear halo bias of mock CMASS galaxies is suggested to have a median (mean) value of $1.7$ ($1.9$) with a systematic discrepancy to the linear prediction at a level of $10\percent$ (see Section~\ref{sec:mocks}),
our constraint of $b_{\mathrm{CMASS}} \approx 1.84 \pm 0.03$ is broadly consistent with that inferred from the N-body simulations ($\approx1.7\sim1.9$) if accounting for the systematic uncertainty.
That is, the consistency in \xicmass\ between the mocks and the observation is largely limited by the systematics, given the precision of the measured \xicmass. 
By adopting the informative prior from the result of the BOSS collaboration \citep{chuang13}, we effectively marginalize the systematic uncertainty of $b_{\mathrm{CMASS}}$ 
in the modelling of \xicamiracmass.
On the other hand, the constraint on $b_{\mathrm{CMASS}}$ becomes significantly more precise, once the \xicmass\ is included in the modelling.
In this case, we observe systematic difference in $b_{\mathrm{CMASS}}$, because the systematic uncertainties are not marginalized over and, hence, are not included.

Meanwhile, the difference between the mock and observed \xicamira\ reflects an offset in the normalization \Arich\ of the \rtm\ relation between the mock and observed cluster samples, additionally to the systematics in the N-body simulations.
For CAMIRA clusters, we evaluate \biash\ following the \citet{tinker10} formula as a function of cluster mass, 
which is mainly determined by the normalization \Arich\ of the \rtm\ relation in the forward-modeling of halo clustering.
Conversely,  the mock clusters are selected by the richness that is assigned by the \rtm\ relation with the normalization \Arich\ calibrated against lensing magnification.
Thus, the difference in the amplitude of \xicamira\ between the mocks and the observation reflects the offset in the absolute mass scale of CAMIRA clusters inferred between lensing magnification and halo clustering.

It is worth mentioning that degeneracy between \Arich\ and $b_{\mathrm{CMASS}}$ seen in the modeling of $\xicamiracmass$ alone (red in Figure~\ref{fig:triangle_plots}) is broken by including the modeling of $\xicamira$ (green contours).
On the other hand, the inclusion of the CMASS auto-correlation (brown in Figure~\ref{fig:triangle_plots}) does not significantly improve the constraint on \Arich\  but only on $b_{\mathrm{CMASS}}$, as opposed to the case of $\xicamira+\xicamiracmass$ (green contours).

\begin{figure}
\centering
\resizebox{0.5\textwidth}{!}{
\includegraphics[scale=1]{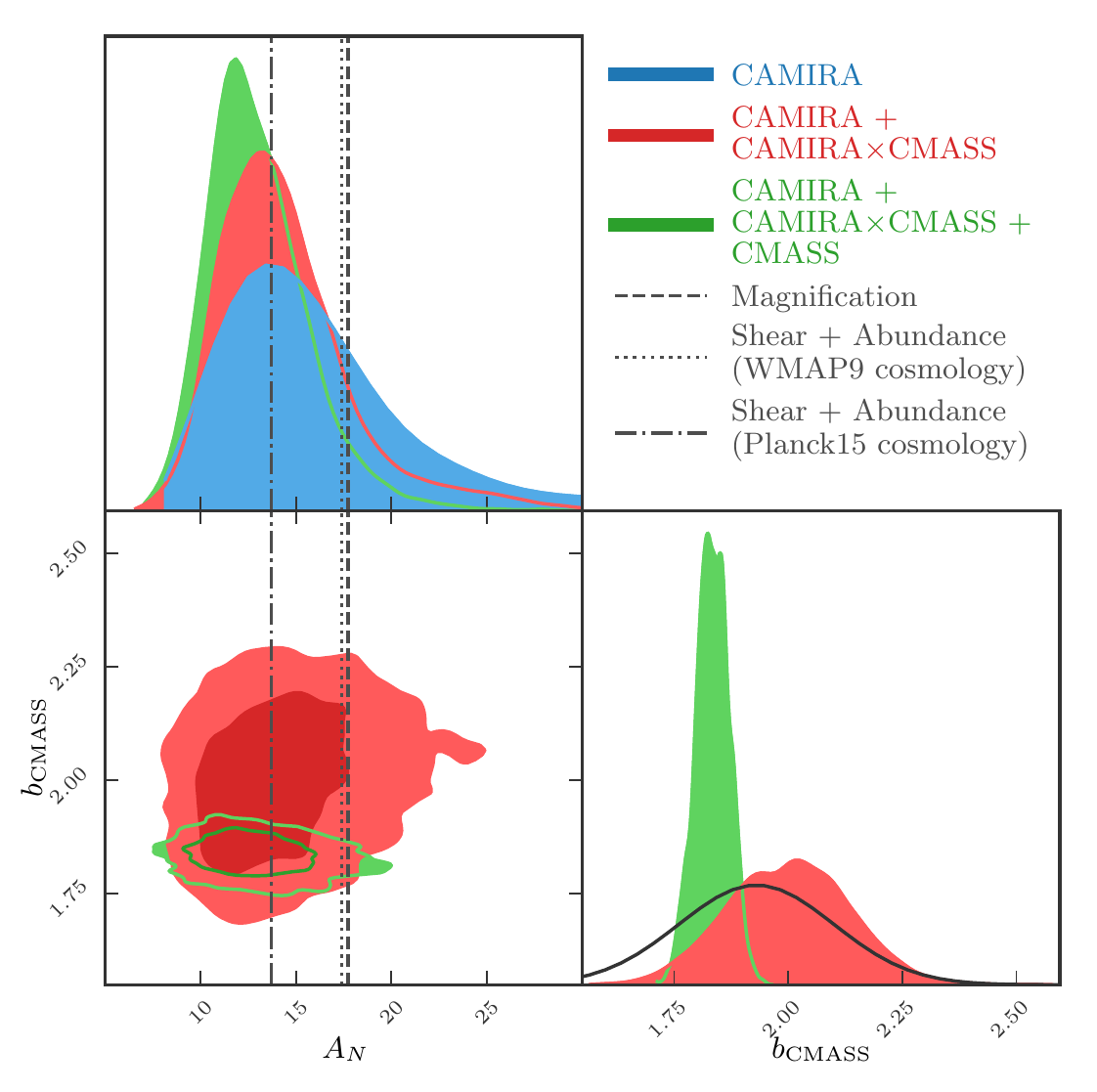}
}
\caption{
Constraints on the normalization \Arich\ of the \rtm\ relation and the linear halo bias $b_{\mathrm{CMASS}}$ of CMASS galaxies.
We show the constraints obtained in the default analysis using the modeling of $\xicamira$, $\xicamira+\xicamiracmass$, and $\xicamira+\xicamiracmass+\xicmass$ in blue, red, and green, respectively.
These modeling are carried out with an informative prior $\mathcal{N}(1.93, 0.17^2)$ on $b_{\mathrm{CMASS}}$, which is adopted from the BAO result \citep{chuang13}, as shown by the black solid line.
This effectively marginalizes over the systematic uncertainty of $b_{\mathrm{CMASS}}$ in the modeling.
Additionally, we show the constraints on \Arich\  that are independently obtained from lensing magnification \citep[][dashed line]{chiu20} and from a joint analysis of cluster abundance and weak shear from \citet{murata19} with the cosmological parameters fixed to those anchored by the $WMAP9$ (dotted line) and the $Planck$ (dotted-dashed line) results.
}
\label{fig:ABplots}
\end{figure}

This picture can be highlighted in Figure~\ref{fig:ABplots}, where we show the constraints of \Arich\ and $b_{\mathrm{CMASS}}$ in the default analysis obtained from the modeling of $\xicamira$ (blue), $\xicamira+\xicamiracmass$ (red), and $\xicamira+\xicamiracmass+\xicmass$ (green).
We find that the uncertainty of \Arich\ is decreased by $\approx30\percent$ by including \xicamiracmass\ into the modeling of \xicamira.
However, including the auto-correlation of the CMASS sample into the joint modeling of $\xicamira+\xicamiracmass$ does not significantly improve the constraint on \Arich.
Based on the  joint modeling of $\xicamira+\xicamiracmass+\xicmass$, we obtain the constraint on \Arich\ as $11.9^{+3.0}_{-1.9}$ with an average precision at a level of $\approx21\percent$.
This constraining power on \Arich\ is comparable to that from lensing magnification alone \citep{chiu20}, which has an uncertainty of $\approx15\percent$ on \Arich.
In Figure~\ref{fig:ABplots}, we additionally show the constraint on \Arich\ inferred from lensing magnification \citep[$\Arich=17.72$, dashed line;][]{chiu20} and from the results of \cite{murata19} using a joint analysis of cluster abundance and weak shear in the cosmology fixed to that anchored by $WMAP9$ ($\Arich=17.40$, dotted line) and \PLANCK\ ($\Arich=13.70$, dotted-dashed line).
We stress that both \cite{chiu20} and \cite{murata19} studied CAMIRA clusters with the same selection (i.e., $0.2\leq\zcl<1.1$ and $\rich\geq15$), which thus enables a direct comparison in this work.
We find that the constraint on \Arich\ using halo clustering are broadly lower than, but statistically consistent with, those inferred from lensing magnification and from a joint analysis of weak shear and cluster abundance at a level of $\lesssim1.9\sigma$, with slight preference for the latter in the \PLANCK\ cosmology.
That is, the clustering-inferred mass scale at fixed richness is higher than those inferred from the independent methods of gravitational lensing and cluster abundance, but not at a statistically significant level ($\lesssim1.9\sigma$).

It is worth noting that the projection effect arising from optical cluster finding algorithms could result in biased lensing signals in the one-halo regime, as suggested by \cite{sunayama2020}.
However, the bias in lensing signals of the one-halo term has a monotonic trend from $\lesssim-5\percent$ to $\lesssim5\percent$ with increasing richness (see Figure~4 in \citealt{sunayama2020}).
Therefore, to first-order approximation, this bias over all clusters in the one-halo term would be averaged out, which is not expected to significantly affect the comparison between the weak-lensing and clustering results.
We will continue to discuss the impact of the projection effect on large-scale clustering in Section~\ref{sec:selection}.

We further note that our constraints on \Arich\ depend on cosmological parameters, especially \seight.
This is because the amplitude of clustering strength is proportional to $\left(\biash\seight\right)^2$, in which \seight\ is fixed to the default value of $0.8$ in this work.
This value is different from that used in the joint analysis of cluster abundance and weak shear in \cite{murata19}, where $\seight=0.82$ ($\seight=0.831$) is used in the \WMAP\ (\PLANCK) cosmology.
Changing \seight\ to $0.82$ ($0.831$) anchored by the \WMAP\ (\PLANCK) cosmology results in a reduction of the halo bias at a level of $1-0.80/0.82\approx2.4\percent$ ($1-0.80/0.831\approx3.7\percent$), implying a mass scale smaller by $5.7\percent$ ($8.7\percent$) at the pivotal mass $\MPIV=10^{14}h^{-1}\Msun$ and the pivotal redshift $\ZPIV=0.6$ assuming the \citet{tinker10} relation.
Since $\Arich\propto{\Mfiveoo}^{-\Brich}$ given an observed richness, the mass scale smaller by $5.7\percent$ ($8.7\percent$) corresponds to an increase in the inferred \Arich\ by $\left(1 - 5.7\percent\right)^{-0.9}\approx5.4\percent$ ($\left(1 - 8.7\percent\right)^{-0.9}\approx8.5\percent$) if changing \seight\ to the value anchored by the \WMAP\ (\PLANCK) cosmology.
That is, our results would be in better agreement with those from \cite{murata19} if accounting for the different \seight\ used in both analysis.
We therefore conclude that the self-calibration of CAMIRA clusters based on halo clustering infers an absolute mass scale that is consistent with those estimated from lensing magnification, weak shear and  cluster abundance (within $\lesssim1.9\sigma$).

%
%

\section{Comments on the selection of the CAMIRA clusters}
\label{sec:selection}

The recent work \cite{sunayama2020} has demonstrated that red-sequence based cluster finders introduce the selection bias to preferentially select galaxy clusters locating at filaments aligning along the line of sight.
This selection bias results in a strong anisotropic pattern in the underlying mass distribution of optically detected clusters on large scale, which ultimately gives boosts to observed lensing signals and the strength of halo clustering compared to the theoretical prediction assuming an isotropic distribution.

In terms of halo clustering, this selection bias leads to a significant quadrupole moment in the 2D correlation function.
In comparison to the halo clustering assuming an isotropic distribution, halos with this selection bias tend to over-cluster (under-cluster) in the direction along (perpendicular to) the line of sight, as seen in Figure~13 of \cite{sunayama2020}.
As a result, there exists a scale-dependent bias in a projected correlation function, in which enhancements of $\approx60\percent$ and $\approx10\percent$ are expected at the projected radius of $R\approx10~h^{-1}\mathrm{Mpc}$ and $R\approx50~h^{-1}\mathrm{Mpc}$, respectively, compared to the case without the selection bias.
A similar picture is implied for lensing signals, where an enhancement up to $\approx20\percent$ is expected in the two-halo term regime.

Since the CAMIRA clusters studied in this work are selected based on a red-sequence finding algorithm, we do expect that such a selection bias exists in our sample.
However, in this work we measured the monopole moment of 3D correlation functions of halos, instead of projected correlation functions as investigated in \cite{sunayama2020}.
Thus, the selection bias is expected to be less significant on our results.
This is because the 3D correlation function is an azimuthal average of the 2D correlation function, such that the effect arising from the quadrupole pattern is significantly alleviated for the 3D correlation function (see a more quantitative
discussion below).


Motivated by \cite{sunayama2020}, we construct a toy model to quantify the effect raised from this selection bias on the 3D correlation function.
We assume that the quadrupole pattern of the 2D correlation function only depends on the cosine angle $\mu_{\mathrm{s}}$, such that the logarithmic ratio of the clustering strength between the observed clusters and the isotropic prediction, $\xi_{\mathrm{obs}}/\xi_{\mathrm{true}}$, follows the relation, $\log_{10}\left(\xi_{\mathrm{obs}}/\xi_{\mathrm{true}}\right) = 0.5\times\mathcal{P}_{2}(\mu_{\mathrm{s}})$, where $\mathcal{P}_{\ell}(\mu_{\mathrm{s}})$ is the $\ell$-th order Legendre polynomial.
We choose the normalization of $0.5$, as the (log-)strength of the anisotropy at $|\mu_{\mathrm{s}}| = 1$, because the second-order Legendre polynomial $\mathcal{P}_{2}$ multiplying this value roughly reproduces the quadrupole pattern of the 2D correlation function in \citet[][see the left panel of their Figure~13]{sunayama2020}.
Then, the integration of $10^{0.5\mathcal{P}_{2}(\mu_{\mathrm{s}})}$ from $\mu_{\mathrm{s}}=0$ to $\mu_{\mathrm{s}}=1$ gives the ratio of $\xi_{\mathrm{obs}}/\xi_{\mathrm{true}} \approx1.15$.
That is, the clustering strength of the 3D correlation function of CAMIRA clusters could be biased high at a level of $\approx15\percent$ compared to that from  an isotropic prediction.
This leads to an overestimated halo bias at a level of $\approx\sqrt{1.15}-1\approx7\percent$.
To first-order approximation, this results in the cluster mass biased high by $\approx17\percent$ at the pivotal mass of $\MPIV=10^{14}h^{-1}\Msun$, assuming that the linear halo bias follows the \citet{tinker10} relation at the pivotal redshift $\ZPIV=0.6$.
Because of $\Arich\propto{\Mfiveoo}^{-\Brich}\approx{\Mfiveoo}^{-0.9}$ given an observed richness, this corresponds to a normalization \Arich\ that is biased low by $\approx13\percent$.
If we change the normalization of $\log\left(\xi_{\mathrm{obs}}/\xi_{\mathrm{true}}\right)$ to $0.3$, then this results in a normalization \Arich\ biased low by $\approx6\percent$.
With this toy model to characterize the effect from the selection bias on the 2D correlation function, the normalization \Arich\ is suggested be biased low by an amount smaller than the current statistical uncertainty, which is at a level of $\approx36\percent$ ($\approx21\percent$) for the modeling of $\xicamira$ ($\xicamira+\xicamiracmass+\xicmass$).
Therefore, we conclude that the selection bias would not significantly alter the interpretation of this work.
However, this selection bias needs to be further quantified in detail if a larger sample of optically detected clusters is studied in future work.

In what follows, we provide two final remarks.
First, in this work we are measuring the 3D correlation function in the redshift-space, in which the photo-\redshift\ uncertainty of CAMIRA clusters would significantly smear out the clustering strength along the line of sight.
This is a distinct difference to \cite{sunayama2020}, where the interpretation is based on projected correlation functions, in which the effects of FoG and RSD are canceled out by integrating along the line of sight.
Moreover, there is no photo-\redshift\ uncertainty present in \cite{sunayama2020}.
The toy model above does not account for the photo-\redshift\ uncertainty, which would mitigate the anisotropic pattern in the 2D correlation function and thus result in an even less impact on \Arich.
Therefore, a quantitative comparison between \cite{sunayama2020} and this work is not trivial.
Second, the interpretation of \cite{sunayama2020} is based on the \texttt{redMaPPer} cluster finder that uses an algorithm different from our \texttt{CAMIRA} finder.
Specifically, the \texttt{redMaPPer} algorithm uses a richness-dependent cutout radius to calculate the observed richness for each cluster in an iterative manner \citep{rykoff14}, while the cutout radius in the \texttt{CAMIRA} algorithm is fixed at a given redshift \citep{oguri14}.
It is also important to note that the \texttt{redMaPPer} cluster finder conducts the global background subtraction in calculating the observed richness, while the local background subtraction is used in the \texttt{CAMIRA} algorithm to account for local variations in large-scale structures around clusters.
These differences introduce a difficulty in quantifying the selection bias of CAMIRA clusters based on the \texttt{redMaPPer} results \citep{murata20}.
A dedicated simulation to quantify the selection bias of \texttt{CAMIRA} clusters is warranted for future work.

%
%

\section{Conclusions}
\label{sec:conclusions}

In this work, we have measured
(1) the auto-correlation function of CAMIRA clusters, \xicamira, with richness $\rich\geq15$ at $0.2\leq\redshift<1.1$, which are constructed using the HSC survey, 
(2) the auto-correlation function of CMASS galaxies, \xicmass, which are spectroscopically observed in the BOSS, and 
(3) the cross-correlation function of these samples, \xicamiracmass.
These correlation functions are measured in redshift space.
Based on these clustering measurements, we carried out a forward-modeling approach to calibrate the \rtm\ relation of CAMIRA clusters, accounting for the effects of the RSD, FoG, and the photo-\redshift\ uncertainty of CAMIRA clusters.
We also take into account the projection effect of the CAMIRA sample on the cluster halo bias.
The modeling is shown to deliver unbiased constraints on the parameters by validation tests against a large set of the mock catalogs, which are carefully constructed from N-body simulations to mimic the observed properties of the CAMIRA and CMASS samples.

We focus on constraining the normalization \Arich\ of the \rtm\ relation, as an absolute calibration of the cluster mass scale, while applying informative priors on other parameters.
In the modeling of $\xicamira$ ($\xicamira+\xicamiracmass$, $\xicamira+\xicamiracmass+\xicmass$), we obtain the constraint of $\Arich = 13.8^{+5.8}_{-4.2}$ ($13.2^{+3.4}_{-2.7}$, $11.9^{+3.0}_{-1.9}$) with an average uncertainty at a level of $36\percent$ ($23\percent$, $21\percent$).
We also carry out a subsample analysis to model the correlation functions in different richness and redshift bins, returning results consistent with the default analysis, given the errorbars.
The self-calibration based on halo clustering alone results in an uncertainty in \Arich\ that is comparable to that independently obtained by lensing magnification \citep{chiu20}, which has an uncertainty at a level of $\approx15\percent$.

We compare the resulting constraints on \Arich\ to those inferred from lensing magnification \citep[][]{chiu20} and from a joint analysis of cluster abundance and weak shear \citep{murata19}.
We find that \Arich\ constrained by halo clustering alone is statistically consistent with the results inferred from those independent methods,
with a preference for a lower \Arich\ (or a higher cluster mass scale) at a level of $\lesssim1.9\sigma$.
Meanwhile, the constraint on the linear halo bias $b_{\mathrm{CMASS}}$ of the CMASS sample is in agreement with the N-body simulations and the observational constraint independently obtained from the 
BOSS
 collaboration \citep{chuang13}, given the uncertainty.

We discuss the effect arising from the selection bias of CAMIRA clusters, in light of the recent work of \cite{sunayama2020}.
We use a simple model to characterize the anisotropic distribution of halo clustering introduced by the selection bias, and assess the potential systematics on  redshift-space correlation functions, as studied in this work.
According to this model, the normalization \Arich\ is suggested to be underestimated by an amount of $\approx13\percent$, to first-order approximation.
This amount is subdominant compared to the current statistical uncertainty at a level between $21\percent$ and $36\percent$, depending on the data sets used in the modeling.
A detailed investigation specifically for the CAMIRA sample is needed if a larger sample is studied in future work.
We also investigate the systematics raised from the cluster random catalog, which is a subdominant factor to our interpretation in this work.

To sum up, we have  shown that the halo clustering of galaxy clusters provides a competitive method in self-calibrating the cluster mass.
By modeling the photo-\redshift\ uncertainty, the redshift-space correlation functions result in a more precise and accurate measurement than the projected or angular correlation functions.
In this work, the clustering-based self-calibration delivers the constraint on the normalization of the \rtm\ relation with a competitive uncertainty of $\approx36\percent$, by only using  $\approx3k$ clusters over a footprint with area of $\approx400~\mathrm{deg}^2$.
Including a spectroscopic sample in a joint analysis of halo clustering 
improves
the uncertainty to $\approx21\percent$.
It is also worth mentioning that clustering analysis is less sensitive to the incompleteness of a tracer sample \citep{guo18}.
Therefore, this paper provides an attractive method of mass calibration for cluster cosmology, paving a way forward with the upcoming large and uniform imaging and spectroscopic surveys (e.g., LSST, DESI, and PFS).

%
%

\section*{Acknowledgments}
\label{sec:acknowledgments}

We thank the anonymous referee for constructive comments, which have led to improvements of this paper.
This work is supported by the Ministry of Science and Technology of Taiwan (grants MOST 106-2628-M-001-003-MY3 and MOST 109-2112-M-001-018-MY3) and by Academia Sinica (grant AS-IA-107-M01).
TO acknowledges support from the Ministry of Science and Technology of Taiwan under Grants No. MOST 106-2119-M-001-031-MY3 and the Career Development Award, Academia Sinina (AS-CDA-108-M02) for the period of 2019-2023.
This work is supported in part by World Premier International Research Center Initiative (WPI Initiative), MEXT, Japan, and JSPS KAKENHI Grant Number JP18K03693.

The Hyper Suprime-Cam (HSC) collaboration includes the astronomical communities of Japan and Taiwan, and Princeton University.  The HSC instrumentation and software were developed by the National Astronomical Observatory of Japan (NAOJ), the Kavli Institute for the Physics and Mathematics of the Universe (Kavli IPMU), the University of Tokyo, the High Energy Accelerator Research Organization (KEK), the Academia Sinica Institute for Astronomy and Astrophysics in Taiwan (ASIAA), and Princeton University.  Funding was contributed by the FIRST program from the Japanese Cabinet Office, the Ministry of Education, Culture, Sports, Science and Technology (MEXT), the Japan Society for the Promotion of Science (JSPS), Japan Science and Technology Agency  (JST), the Toray Science  Foundation, NAOJ, Kavli IPMU, KEK, ASIAA, and Princeton University.

This paper makes use of software developed for the Legacy Survey of Space and Time (LSST) carried out by the Vera C. Rubin Observatory. We thank the LSST Project for making their code available as free software at  http://dm.lsst.org

This paper is based in part on data collected at the Subaru Telescope and retrieved from the HSC data archive system, which is operated by Subaru Telescope and Astronomy Data Center (ADC) at NAOJ. Data analysis was in part carried out with the cooperation of Center for Computational Astrophysics (CfCA), NAOJ.

The Pan-STARRS1 Surveys (PS1) and the PS1 public science archive have been made possible through contributions by the Institute for Astronomy, the University of Hawaii, the Pan-STARRS Project Office, the Max Planck Society and its participating institutes, the Max Planck Institute for Astronomy, Heidelberg, and the Max Planck Institute for Extraterrestrial Physics, Garching, The Johns Hopkins University, Durham University, the University of Edinburgh, the Queen's University Belfast, the Harvard-Smithsonian Center for Astrophysics, the Las Cumbres Observatory Global Telescope Network Incorporated, the National Central University of Taiwan, the Space Telescope Science Institute, the National Aeronautics and Space Administration under grant No. NNX08AR22G issued through the Planetary Science Division of the NASA Science Mission Directorate, the National Science Foundation grant No. AST-1238877, the University of Maryland, Eotvos Lorand University (ELTE), the Los Alamos National Laboratory, and the Gordon and Betty Moore Foundation.

Funding for SDSS-III has been provided by the Alfred P. Sloan Foundation, the Participating Institutions, the National Science Foundation, and the U.S. Department of Energy Office of Science. The SDSS-III web site is \url{http://www.sdss3.org}.

SDSS-III is managed by the Astrophysical Research Consortium for the Participating Institutions of the SDSS-III Collaboration including the University of Arizona, the Brazilian Participation Group, Brookhaven National Laboratory, Carnegie Mellon University, University of Florida, the French Participation Group, the German Participation Group, Harvard University, the Instituto de Astrofisica de Canarias, the Michigan State/Notre Dame/JINA Participation Group, Johns Hopkins University, Lawrence Berkeley National Laboratory, Max Planck Institute for Astrophysics, Max Planck Institute for Extraterrestrial Physics, New Mexico State University, New York University, Ohio State University, Pennsylvania State University, University of Portsmouth, Princeton University, the Spanish Participation Group, University of Tokyo, University of Utah, Vanderbilt University, University of Virginia, University of Washington, and Yale University.

This work is possible because of the efforts in the LSST \citep{juric17,ivezic19} and PS1 \citep{chambers16, schlafly12, tonry12, magnier13}, and in the HSC \citep{aihara18a} developments including the deep imaging of the COSMOS field \citep{tanaka17}, the on-site quality-assurance system \citep{furusawa18}, the Hyper Suprime-Cam \citep{miyazaki15, miyazaki18, komiyama18}, the design of the filters \citep{kawanomoto18},  the data pipeline \citep{bosch18}, the design of bright-star masks \citep{coupon18}, the characterization of the photometry by the code \texttt{Synpipe} \citep{huang18}, the photometric redshift estimation \citep{tanaka18}, the shear calibration \citep{mandelbaum18}, and the public data releases \citep{aihara18b, aihara19}.

This work made use of the IPython package \citep{PER-GRA:2007}, SciPy \citep{jones_scipy_2001}, TOPCAT, an interactive graphical viewer and editor for tabular data \citep{2005ASPC..347...29T}, matplotlib, a Python library for publication quality graphics \citep{Hunter:2007}, Astropy, a community-developed core Python package for Astronomy \citep{2013A&A...558A..33A}, NumPy \citep{van2011numpy}. 
This work made use of \citet{bocquet16b} and \cite{hinton2016} for producing the corner plots for the parameter constraints.

%
%

\appendix

\section{The parameter constraints}
\label{sec:params_without_priors}

\begin{figure*}
\centering
\resizebox{\textwidth}{!}{
\includegraphics[scale=1]{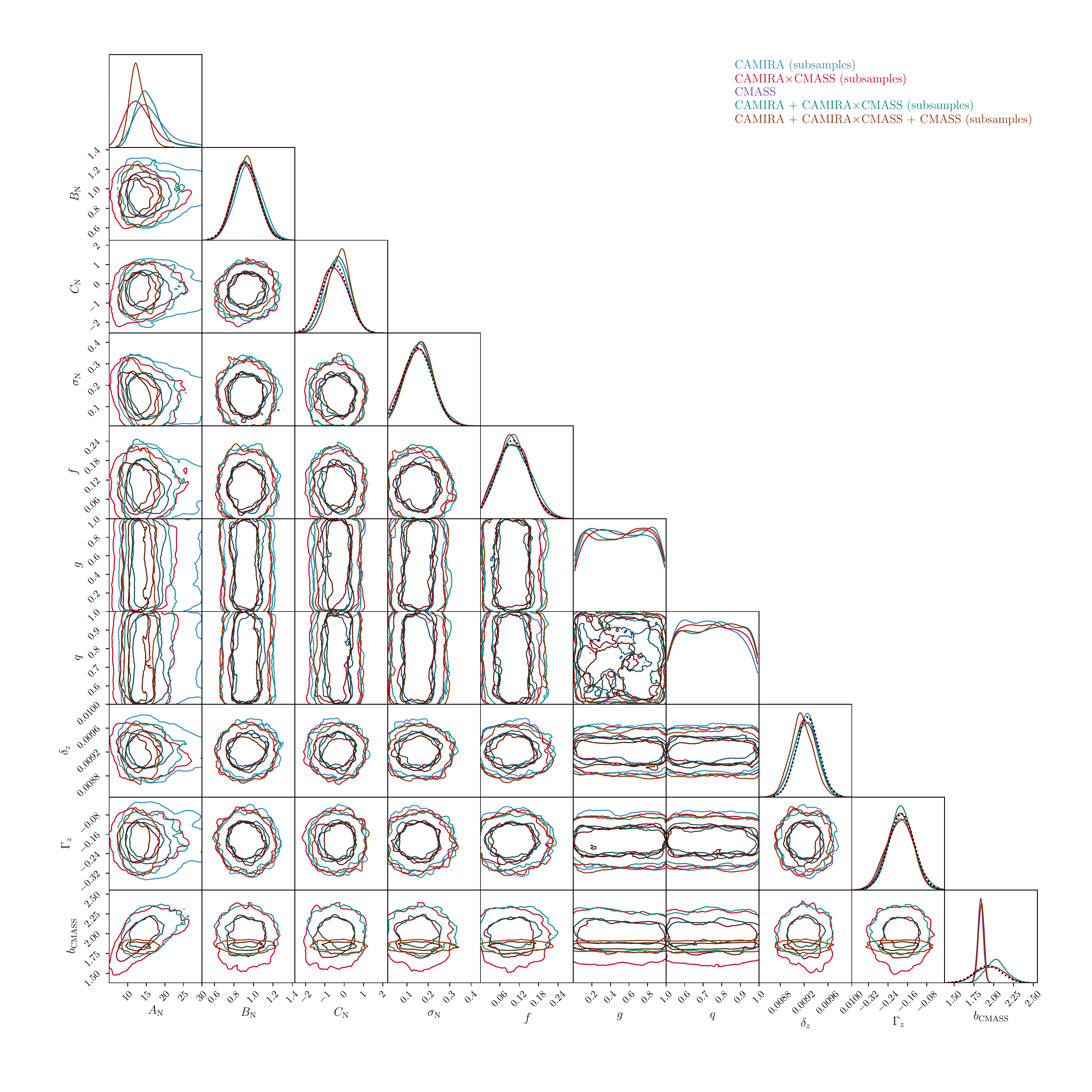}
}\vspace{-1.2cm}
\caption{
Parameter constraints obtained in the subsample analysis.
This plot is generated in the same configuration as in Figure~\ref{fig:triangle_plots_3dmocktests}.
}
\label{fig:triangle_plots_subsamples}
\end{figure*}

\begin{figure*}
\centering
\resizebox{\textwidth}{!}{
\includegraphics[scale=1]{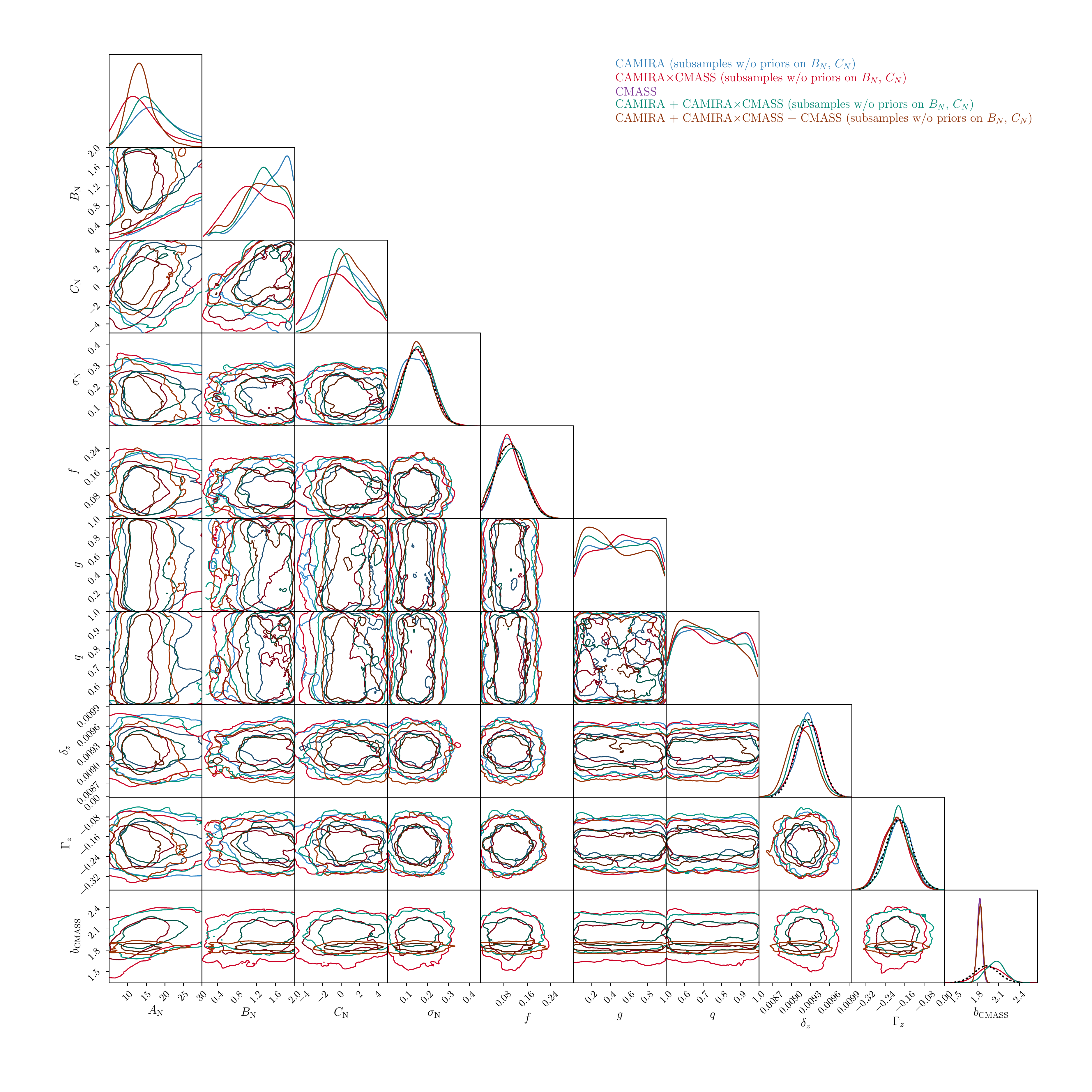}
}\vspace{-1.2cm}
\caption{
Parameter constraints obtained in the subsample analysis without the Gaussian priors on the mass- and redshift-trend power-law indices of the \rtm\ relation.
This plot is generated in the same configuration as in Figure~\ref{fig:triangle_plots_3dmocktests}.
}
\label{fig:triangle_plots_without_prior}
\end{figure*}

We repeat the analysis in our subsample analysis, in which we bin our
cluster sample based on observed richness and redshift with four
subsamples in total (see Section~\ref{sec:measurements_of_correlations}).
In this section, we show the parameter constraints obtained in the subsample analysis in Figure~\ref{fig:triangle_plots_subsamples} and those without the Gaussian priors (see Section~\ref{sec:modeling}) on \Brich\ and \Crich\ in the subsample analysis in Figure~\ref{fig:triangle_plots_without_prior}.

%

%
%

\section{Systematics due to the cluster random catalog}
\label{sec:sys}

\begin{figure}
\centering
\resizebox{0.5\textwidth}{!}{
\includegraphics[scale=1]{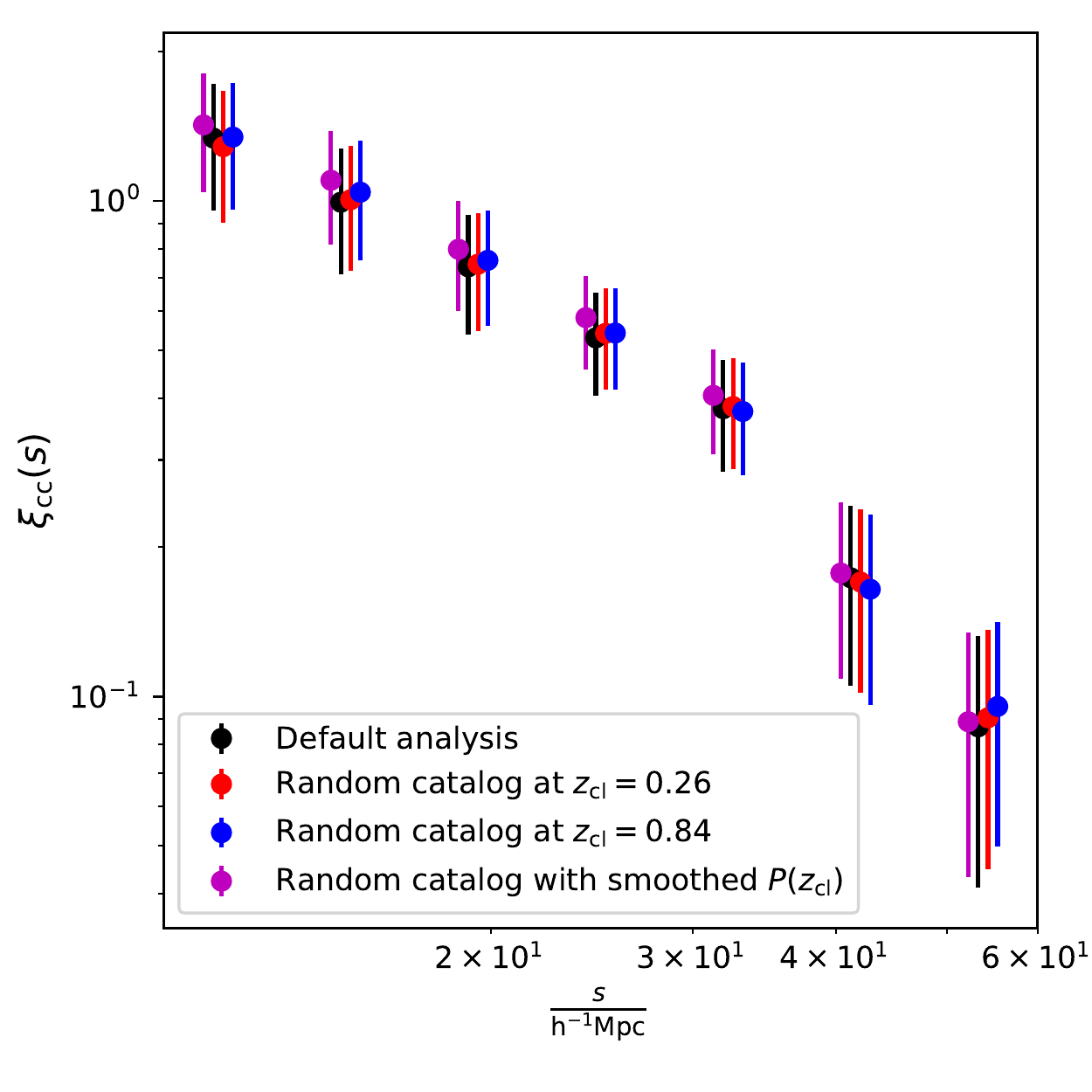}
}
\caption{
Comparison of auto-correlation functions \xicamira\ of CAMIRA clusters with $\rich\geq15$ at $0.2\leq\redshift<1.1$ among different random catalogs.
The black circles are the result of the default analysis, with the random catalog generated at the redshift of $0.56$.
The results using the cluster random catalogs generated at redshift of $0.26$ and $0.84$ are shown as the red and blue circles, respectively.
The resulting \xicamira\ 
using the random catalog with the smoothed redshift distribution of observed CAMIRA clusters is indicated by the purple circles (see Section~\ref{sec:sys} for more details). 
The difference among these results is negligible compared to the size of current errorbars.
}
\label{fig:sys_random}
\end{figure}

In this section, we assess the systematics introduced by the random catalog of the CAMIRA clusters.
The construction of the random catalog for the CAMIRA clusters consists of two phases: the randomization of (1) the angular distribution and (2) the redshift distribution.
For the former, the default analysis is done using the random cluster catalog based on the aperture with the fixed angular size at the redshift of $0.56$, approximately the median redshift of the CAMIRA clusters.
To assess the systematics from the angular randomization of the random catalog at the fixed redshift, we re-measured \xicamira\ of all the CAMIRA clusters with the random cluster catalogs produced at redshift of $0.26$ and $0.89$, as shown by the red and blue circles in Figure~\ref{fig:sys_random}, respectively.
As seen in Figure~\ref{fig:sys_random}, the resulting \xicamira\ is insensitive to the choice of redshift where the random catalog is produced, given the errorbars.
We therefore conclude that the chosen redshift in generating the  angular distribution of the cluster random catalog is a subdominant factor in our analysis.

For randomization of the redshift distribution, we use the redshifts ``shuffled''  from the observed clusters (i.e., bootstrapping the redshift estimates from the CAMIRA cluster catalog).
We also assess the systematics raised from the random distribution of the redshifts in the similar way as done in Section~6 of \cite{ross12}.
Specifically, we alternatively assign the redshift estimate to each point in the random catalog following the observed redshift distribution of the CAMIRA clusters after the smoothing using a Gaussian kernel.
We use a Gaussian kernel with a dispersion of $0.009$, which is the observed dispersion in the photo-\redshift\ uncertainty (see Section~\ref{sec:modeling}), to convolve the redshift distribution of the CAMIRA clusters with $\rich\geq15$ derived using a redshift step of $0.002$.
The size of the redshift step is chosen in order to have enough sampling to resolve one interval of the redshift dispersion.
The resulting \xicamira\
is shown by the blue circles in Figure~\ref{fig:sys_random}.
Given the errorbars, the difference to the default analysis (black circles) is negligible.
To sum up, the interpretation of this work thus remains intact from the systematics introduced from the cluster random catalogs.

\section*{Data Availability}
The data underlying this article were accessed from the Hyper Suprime-Cam collaboration and the public data release of the Baryon Oscillation Spectroscopic Survey. The derived data generated in this research will be shared on reasonable request to the corresponding author.

\bibliographystyle{mn2e}
\bibliography{literature}

\end{document}

%% file: authorlist.tex
\newcommand{\ASIAA}{$^{1}$}
\newcommand{\IPMU}{$^{2}$}
\newcommand{\UTOKYO}{$^{3}$}
\newcommand{\RESCUE}{$^{4}$}

\author[Chiu et al.]{
\parbox{\textwidth}{
\LARGE
\vspace{-0.8cm}
I-Non Chiu\ASIAA,
Teppei Okumura\ASIAA$^{,}$\IPMU,
Masamune Oguri\IPMU$^{,}$\UTOKYO$^{,}$\RESCUE,
Aniket Agrawal\ASIAA,\\
Keiichi Umetsu\ASIAA,
and
Yen-Ting Lin\ASIAA
}
\vspace{0.4cm}
\\
\parbox{\textwidth}{
\ASIAA\ Academia Sinica Institute of Astronomy and Astrophysics (ASIAA), 11F of AS/NTU Astronomy-Mathematics Building, No.1, Sec. 4, Roosevelt Rd, Taipei 10617, Taiwan\\
\IPMU\ Kavli Institute for the Physics and Mathematics of the Universe (WPI), The University of Tokyo Institutes for Advanced Study (UTIAS), The University of Tokyo, 5-1-5 Kashiwanoha, Kashiwa-shi, Chiba, 277-8583, Japan\\
\UTOKYO\ Department of Physics, University of Tokyo, 7-3-1 Hongo, Bunkyo-ku, Tokyo 113-0033 Japan\\
\RESCUE\ Research Center for the Early Universe, University of Tokyo, Tokyo 113-0033, Japan
}
}